\pdfoutput=1
\documentclass[11pt]{article}

\usepackage[margin=0.9in]{geometry}
\usepackage{amsmath,amssymb,amsthm,mathtools,bm}
\usepackage{graphicx}
\usepackage{booktabs,array}
\usepackage{float}
\usepackage{placeins}
\usepackage{enumerate}
\usepackage{algorithm}
\usepackage{microtype}
\usepackage[round,authoryear]{natbib}
\usepackage[
  colorlinks=true,
  linkcolor=blue,
  citecolor=blue,
  urlcolor=blue
]{hyperref}
\usepackage{doi}
\usepackage{authblk}

\theoremstyle{plain}
\newtheorem{theorem}{Theorem}
\newtheorem{proposition}{Proposition}
\newtheorem{lemma}{Lemma}
\newtheorem{corollary}{Corollary}
\theoremstyle{definition}
\newtheorem{definition}{Definition}
\newtheorem{assumption}{Assumption}
\theoremstyle{remark}
\newtheorem{remark}{Remark}
\newtheorem{example}{Example}

\newcommand{\R}{\mathbb{R}}
\newcommand{\bbP}{\mathbb{P}}
\newcommand{\bbE}{\mathbb{E}}
\newcommand{\cH}{\mathcal{H}}
\newcommand{\cC}{\mathcal{C}}
\newcommand{\cO}{\mathcal{O}}
\newcommand{\cY}{\mathcal{Y}}
\newcommand{\trans}{\mathsf{T}}
\newcommand{\Var}{\operatorname{Var}}
\newcommand{\Cov}{\operatorname{Cov}}
\newcommand{\MMD}{\operatorname{MMD}}
\newcommand{\expit}{\operatorname{expit}}
\newcommand{\tablebodysetup}{\footnotesize\renewcommand{\arraystretch}{1.08}}
\DeclareMathOperator*{\argmax}{arg\,max}

\newcommand{\suppcounter}[1]{%
  \setcounter{#1}{0}%
  \expandafter\renewcommand\csname the#1\endcsname{S\arabic{#1}}%
  \expandafter\def\csname theH#1\endcsname{S\arabic{#1}}}

\allowdisplaybreaks
\title{On Ignorability Beyond Missing at Random:\\
Completion-Invariant Missingness for Nonmonotone Missing Data}
\author[1,$\ast$]{Kosuke Morikawa}
\author[2]{Yutaka Kano}
\author[3]{Shonosuke Sugasawa}
\affil[1]{Department of Statistics, Iowa State University, Ames, Iowa, USA}
\affil[2]{Faculty of Culture and Information Science, Doshisha University, Kyotanabe, Kyoto, Japan}
\affil[3]{Faculty of Economics, Keio University, Tokyo, Japan}
\affil[$\ast$]{Corresponding author. Email: \texttt{morikawa@iastate.edu}}
\date{}
\hypersetup{
  pdftitle={On Ignorability Beyond Missing at Random: Completion-Invariant Missingness for Nonmonotone Missing Data},
  pdfauthor={Kosuke Morikawa, Yutaka Kano, Shonosuke Sugasawa}
}

\begin{document}
\maketitle

\begin{abstract}
The missing-at-random assumption underlies most analyses of incomplete data. Under nonmonotone missingness, it is especially stringent, imposing pattern-specific restrictions with different conditioning sets. A strictly weaker normalization condition is known to justify maximizing the observed-data likelihood. We show that every law satisfying this condition has a unique missing-at-random counterpart sharing its observed-data and complete-data distributions. Observed-data procedures developed under missing at random for identified functionals, including doubly robust estimators, therefore remain valid, with the same semiparametric efficiency bound. We then prove that the condition holds if and only if filling in the missing values from their conditional distribution under the true full model preserves the complete-data distribution, hence the name completion-invariant missingness. We use a characteristic kernel to compare the completed distribution with the fitted model and assess the joint compatibility of the specified full model and completion-invariant missingness. We derive the asymptotic null distribution with estimated parameters and establish bootstrap and multiplier calibration. Since a parametric full model alone need not identify completion-invariant missingness, we give identification conditions for Gaussian, Student $t$ and finite mixture families. Simulations assess sensitivity to departures from each assumption, and applications to clinical laboratory measurements and childhood growth illustrate the diagnostic.
\end{abstract}

\noindent\textbf{Keywords:} completion-invariant missingness, direct likelihood, kernel discrepancy test, missing at random, nonmonotone missing data, semiparametric efficiency

\section{Introduction}\label{sec:intro}

Missing at random (MAR) is central to the analysis of incomplete data. Under MAR and parameter distinctness, likelihood-based inference for a complete-data model can proceed without explicitly modelling the missing-data mechanism \citep{Rubin1976}, which permits estimation, imputation and uncertainty quantification from the substantive model despite incomplete observations. MAR is not generally verifiable from the observed distribution: its testability depends on the observation structure \citep{GillVanDerLaanRobins1997}, and an unrestricted complete-data law can admit observationally equivalent missing-not-at-random (MNAR) explanations \citep{MolenberghsEtAl2008}. The difficulty is pronounced under nonmonotone missingness, where MAR imposes a separate restriction for each pattern with its own conditioning set, and, outside the randomized monotone missingness class described below, mechanisms satisfying all of them are hard to motivate \citep{RobinsGill1997}.

Several approaches to nonmonotone missing data have been proposed. Randomized monotone missingness \citep{RobinsGill1997} generates MAR mechanisms by sequentially choosing which variable to observe next, or whether to stop, using only previously observed values; because the observation order can differ across units, it yields nonmonotone patterns while remaining a proper subclass of MAR. Other approaches assume MAR and proceed by likelihood or imputation \citep{Schafer1997} or by weighted estimating equations \citep{SunTchetgen2018}, or impose MNAR restrictions through discrete-choice and self-censoring models \citep{TchetgenWangSun2018,MalinskyEtAl2022}, pattern mixtures \citep{Little1993} or pattern graphs \citep{Chen2022PatternGraphs}. Existing diagnostics address MCAR \citep{Little1988}, MAR+, a testable strengthening of MAR \citep{PotthoffEtAl2006}, or missing always at random, the everywhere formulation used here, under additional assumptions \citep{BojinovEtAl2020}. Other diagnostics rely on instrumental variables \citep{Breunig2019} or parametric response models \citep{WangEtAl2023}. These approaches thus either strengthen MAR, assume it, or replace it with MNAR restrictions.

This paper studies a condition strictly weaker than MAR. \citet{Kano2014} showed that direct-likelihood estimation, the maximization of the observed-data likelihood, remains consistent under a normalization condition that requires only that the pattern probabilities, each conditioned on the values observed under its own pattern, sum to one. We call this condition \emph{completion-invariant missingness} (CIM), formally defined in Definition~\ref{def:cim}. CIM is strictly weaker than MAR, so consistency of the substantive parameter estimates does not require MAR. We ask what this weakening costs relative to MAR-based inference, in particular whether the semiparametric efficiency bound rises and by how much, what CIM means, and how to assess its compatibility with a specified model for the joint distribution of the complete-data vector.

Our first result is observed-data equivalence, which shows that the weakening costs nothing for functionals of the complete-data marginal. For finite-support complete data with an unrestricted complete-data law, \citet[Theorem~2.1]{RobinsGill1997}, following \citet{GillVanDerLaanRobins1997}, obtained an MAR representation of every observed-data law by selecting a generally different complete-data marginal. \citet{MolenberghsEtAl2008} instead used the complete-data law of a given model, and \citet{Kano2014} showed that this counterpart is MAR if and only if CIM holds. We prove that CIM exactly characterizes when no replacement of the complete-data marginal is needed: an MAR law with the same observed-data and complete-data distributions exists if and only if CIM holds, and it is then unique, so that for any full model the attainable pairs of observed-data law and complete-data marginal coincide with those under MAR. Every observed-data statistic therefore has the same sampling distribution under such a law and under its MAR counterpart, and for identified, pathwise differentiable functionals of the complete-data marginal the efficient influence function and the semiparametric efficiency bound are unchanged. Therefore, augmented inverse-probability-weighted and doubly robust procedures developed under MAR \citep{RobinsRotnitzkyZhao1994,BangRobins2005} retain their validity, with the same standard errors and efficiency benchmark. This inferential transfer requires neither a parametric specification of the actual missing-data mechanism nor an assumption of parameter distinctness.

Our second result characterizes CIM through conditional completion. This operation retains each unit's observed values and draws its missing values from their conditional distribution under a full model. CIM holds if and only if completion under the true full model preserves the complete-data distribution. To assess whether the observed data are jointly compatible with CIM and a specified full model, we compare the completed distribution with the fitted model using a characteristic kernel \citep{SriperumbudurEtAl2010,Gretton2012}. The null distribution of the fitted kernel statistic is derived accounting for direct-likelihood estimation, centred refitted-bootstrap and influence-function multiplier calibrations are shown to be valid, and the test is consistent against fixed alternatives that remain separated after fitting.

Our third result concerns identification. A parametric complete-data family alone need not determine whether CIM holds: a non-CIM Gaussian law and a missing-completely-at-random (MCAR) Gaussian law can share the same observed-data distribution, as Supplementary Example~\ref{ex:gaussian-observational-equivalence} shows. We give sufficient conditions that exclude such observationally equivalent CIM explanations, based either on a tail condition on the complete-case probability under a Gaussian full model, or on monotonicity of the observation probability in a variable exclusive to one pattern whose conditional distribution is Gaussian or Student $t$; the latter rests on a conditional-independence property that CIM shares with MAR. Existing MNAR identification results suffice whenever they identify the complete-data marginal over the same comparison class \citep{MiaoDingGeng2016,MiaoLiuTchetgenGeng2019}. For arbitrary nonmonotone patterns, testability requires restrictions on the complete-data law, the mechanism or the available variables; ours fall on the complete-data law and, for identification, on the observation probabilities.

Section~\ref{sec:setup} defines CIM and relates it to MAR and to direct-likelihood estimation. Section~\ref{sec:estimation} establishes observed-data equivalence and efficiency invariance. Section~\ref{sec:tests} develops the completion characterization and the fitted-kernel test with its calibration. Section~\ref{sec:identification} gives the identification conditions. Section~\ref{sec:simulation} examines inferential invariance, null calibration and power by simulation, and Section~\ref{sec:real-data} illustrates the diagnostic with clinical laboratory measurements and childhood growth data. Section~\ref{sec:conclusion} concludes. Proofs and computational details are in the Supplementary Material.

\section{Framework for Completion-invariant Missingness}\label{sec:setup}

Let $Y=(Y_1,\ldots,Y_p)$ have Borel support $\cY\subseteq\R^p$, full model $f(y;\theta)$, $\theta\in\Theta$, and true marginal law $P_Y=P_{Y,\theta_0}$. The \emph{full model} specifies the joint distribution of $Y$ only; the missing-data mechanism, the conditional distribution of $R$ given $Y$, is separate. The finite pattern set is $\mathcal R\subseteq\{0,1\}^p$; $Y^{(r)}$ and $Y^{(-r)}$ denote coordinates observed and missing under $r$. An observed row is $O=(R,Y^{(R)})$, and $\cO_n=\{O_i:i=1,\ldots,n\}$. When present, $\mathbf1$ and $\mathbf0$ denote the complete and all-missing patterns. Observed rows are independent and identically distributed.

Densities use a product measure $\mu$ with coordinate factors $\mu^{(r)}$ and $\mu^{(-r)}$. Write $\rho_r=\bbP(R=r)$ and distinguish the observed-pattern density $f_r(u\mid R=r)$ from the full-model marginal $f_r(u;\theta)=\int f(u,v;\theta)\,d\mu^{(-r)}(v)$. Any coherent joint full model with evaluable marginals and conditional completion distributions is allowed. Define
\begin{equation}
\pi_r(u)=\bbP(R=r\mid Y^{(r)}=u)
=\rho_r\frac{f_r(u\mid R=r)}{f_r(u;\theta_0)}.
\label{eq:pi}
\end{equation}
The first identity defines $\pi_r$ from the joint law; the second uses correct full-model specification and Bayes' formula. MAR means $\bbP(R=r\mid Y=y)=\pi_r(y^{(r)})$ for every $r$, $P_Y$-almost surely. This is the unit-level, almost-sure formulation of \emph{everywhere MAR} \citep{SeamanEtAl2013}, applying to all patterns and observed values, rather than just the realized sample; an MAR model imposes it at every law.

We now introduce a weaker condition on the missing-data mechanism, replacing the patternwise MAR restrictions by a single normalization identity.

\begin{definition}[Completion-invariant missingness]\label{def:cim}
CIM holds if
\begin{equation}
\sum_{r\in\mathcal R}\pi_r(y^{(r)})=1
\qquad P_{Y,\theta_0}\text{-a.s.}
\label{eq:cim}
\end{equation}
\end{definition}

\subsection{Relation to MAR}\label{sec:relation-mar}
MAR implies CIM by summing its conditional pattern probabilities. To see the weaker restriction, set $F_r(Y)=\pi_r(Y^{(r)})-\bbP(R=r\mid Y)$ and
\begin{equation}
F(Y)=\sum_r F_r(Y)=\sum_r\pi_r(Y^{(r)})-1.
\label{eq:F}
\end{equation}
MAR sets every $F_r$ to zero; CIM requires only $F=0$. Conversely, patternwise conditionally mean-zero perturbations of an MAR mechanism preserve CIM if they sum to zero and leave nonnegative pattern probabilities; both the observed law and complete-data marginal are preserved. With only the complete pattern and one incomplete pattern, CIM and MAR coincide \citep[Proposition~4.2]{Kano2014}. For example, with patterns $11$ and $10$, CIM gives $\pi_{11}(y_1,y_2)=1-\pi_{10}(y_1)=\bbP(R=11\mid Y_1=y_1)$, which implies MAR. Even monotone missingness allows CIM--MNAR mechanisms once a second incomplete pattern is present.

\begin{proposition}[Three monotone patterns]\label{prop:three-pattern-counterexample}
Let $Y=(Y_1,Y_2,Y_3)$ and let the possible patterns be
$111$, $110$ and $100$.  Choose positive constants $\rho_{111},\rho_{110},\rho_{100}$ with
$\rho_{111}+\rho_{110}+\rho_{100}=1$ and a function $H(Y)$, not almost surely zero, satisfying $|H(Y)|\leq1$ and $\bbE(H(Y)\mid Y_1,Y_2)=0$.
For $0<\epsilon<\min(\rho_{110},\rho_{100})$, define $\bbP(R=111\mid Y)=\rho_{111}$, $\bbP(R=110\mid Y)=\rho_{110}+\epsilon H(Y)$ and $\bbP(R=100\mid Y)=\rho_{100}-\epsilon H(Y)$.
Then CIM holds.  
If $H(Y)$ depends on $Y_3$, the mechanism is MNAR.
Moreover, its observed-data distribution is identical to that of the MCAR
mechanism with pattern probabilities $(\rho_{111},\rho_{110},\rho_{100})$.
\end{proposition}

For a multivariate Gaussian model, take $H(Y)=\tanh(e_3)$, where $e_3$ is the standardized residual of $Y_3$ given $(Y_1,Y_2)$. Independence and symmetry of $e_3$ give $\bbE(H\mid Y_1,Y_2)=0$.

CIM does not imply $\mathcal L(Y^{(-r)}\mid Y^{(r)},R=r)=\mathcal L(Y^{(-r)}\mid Y^{(r)})$: the normalization identity can hold even when, within a missingness pattern, the conditional distribution of the missing values given the observed values differs from the full-model conditional. Individual imputations and pattern-specific targets require the corresponding patternwise ignorability identity. In Proposition~\ref{prop:three-pattern-counterexample} with $Y_1,Y_2,Y_3$ independent and uniform on $\{-1,1\}$, $\rho_{111}=\rho_{110}=\rho_{100}=1/3$, $H(Y)=Y_3$ and $\epsilon=1/6$, CIM holds, yet $\bbP(Y_3=1\mid R=110)=3/4$ whereas the full-model conditional gives $1/2$. CIM does retain a conditional-independence property also shared by MAR for variables exclusive to one pattern (see Proposition~\ref{prop:exclusive-variable}).

\subsection{Direct-likelihood estimation}
Omitting the missing-data mechanism gives the \emph{direct-likelihood} criterion and estimator
\begin{align}
\ell_n^{\mathrm{obs}}(\theta)
&=\sum_{i=1}^n\log f_{R_i}(Y_i^{(R_i)};\theta),
\qquad \widehat\theta\in\argmax_{\theta\in\Theta}\ell_n^{\mathrm{obs}}(\theta),
\label{eq:observed-mle}\\
\bar\ell(\theta)
&=\bbE\{\log f_R(Y^{(R)};\theta)\}.
\label{eq:population-objective}
\end{align}
Under correct full-model specification and CIM, together with suitable support, observed-data identification and uniform convergence conditions, $\widehat\theta\overset p\longrightarrow\theta_0$ \citep{Kano2014}. Supplementary Lemma~\ref{lem:consistency} states these conditions. Identification from complete data alone does not ensure identification after missingness.

\section{Observed-data equivalence and inference}\label{sec:estimation}\label{sec:estimation-transfer}

Let $\mathcal M$ be any model for the complete-data marginal, including the parametric family above. For a joint law $P_{Y,R}$ write $P_Y$ and $P_O$ for its complete-data and observed-data marginals, using fixed measurable conditional-probability versions, which exist on the Borel space. Probabilities and expectations without subscripts refer to this law. Let $\mathcal P_{\mathrm{CIM}}(\mathcal M)$ and $\mathcal P_{\mathrm{MAR}}(\mathcal M)$ contain the joint laws with $P_Y\in\mathcal M$ satisfying CIM and MAR, respectively, with no further mechanism restrictions.

\begin{definition}[Counterpart construction]\label{def:counterpart}
For a joint law $P_{Y,R}$ of $(Y,R)$, define its \emph{counterpart} $(P_{Y,R})^{\mathrm{cp}}$ by
\begin{equation}
(P_{Y,R})^{\mathrm{cp}}(Y\in dy,R=r)
=\pi_r(y^{(r)})\,P_Y(dy),\qquad r\in\mathcal R.
\label{eq:kano-counterpart-law}
\end{equation}
It is called an MAR counterpart when its missing-data mechanism is MAR.
\end{definition}

Write $P_Y^{\mathrm{cp}}$ and $P_O^{\mathrm{cp}}$ for its two marginals. The construction above is the counterpart of \citet{MolenberghsEtAl2008}, written in terms of probability measures: it is built from the original complete-data marginal $P_Y$ and always preserves $P_O$, but it need not preserve $P_Y$ as its own marginal. In the finite-support setting with an unrestricted complete-data law, \citet[Theorem~2.1]{RobinsGill1997}, following \citet{GillVanDerLaanRobins1997}, likewise preserve the observed-data law, but select a generally different complete-data law so that an MAR representation exists for every observed-data law. \citet[Proposition~4.4]{Kano2014} established that the construction based on the original $P_Y$ is MAR if and only if the original joint law satisfies CIM, yielding the equivalence of (i) and (ii) below. Theorem~\ref{thm:estimation-transfer} formulates these relations for general Borel laws, adds the equivalence of (i) and (iii), marginal preservation, and uses it to establish the existence and uniqueness of the MAR counterpart with both marginals fixed and the equality of the attainable pairs $(P_O,P_Y)$; these results yield the inferential transfer and efficiency statements below and connect them to the completion diagnostic of Section~\ref{sec:tests}.

\begin{theorem}[Marginal preservation and observed-data model equivalence]\label{thm:estimation-transfer}
For every joint law $P_{Y,R}$, its counterpart in
Definition~\ref{def:counterpart} is a probability law and satisfies
$P_O^{\mathrm{cp}}=P_O$. The following are equivalent:
\begin{enumerate}[(iii)]
\item[(i)] $P_{Y,R}$ satisfies CIM.
\item[(ii)] $(P_{Y,R})^{\mathrm{cp}}$ is MAR.
\item[(iii)] $P_Y^{\mathrm{cp}}=P_Y$.
\end{enumerate}
Equivalently, an MAR law with the same observed-data and complete-data
marginals as $P_{Y,R}$ exists if and only if conditions (i)--(iii) hold.
When it exists, it is unique and equals $(P_{Y,R})^{\mathrm{cp}}$.
Consequently, for every full model $\mathcal M$,
\begin{align}
&\left\{
\bigl(P_O,P_Y\bigr):
P_{Y,R}\in\mathcal P_{\mathrm{CIM}}(\mathcal M)
\right\}=
\left\{
\bigl(P_O,P_Y\bigr):
P_{Y,R}\in\mathcal P_{\mathrm{MAR}}(\mathcal M)
\right\}.
\label{eq:parameter-model-equivalence}
\end{align}
\end{theorem}

\begin{figure}[H]
\centering
\includegraphics[width=95mm]{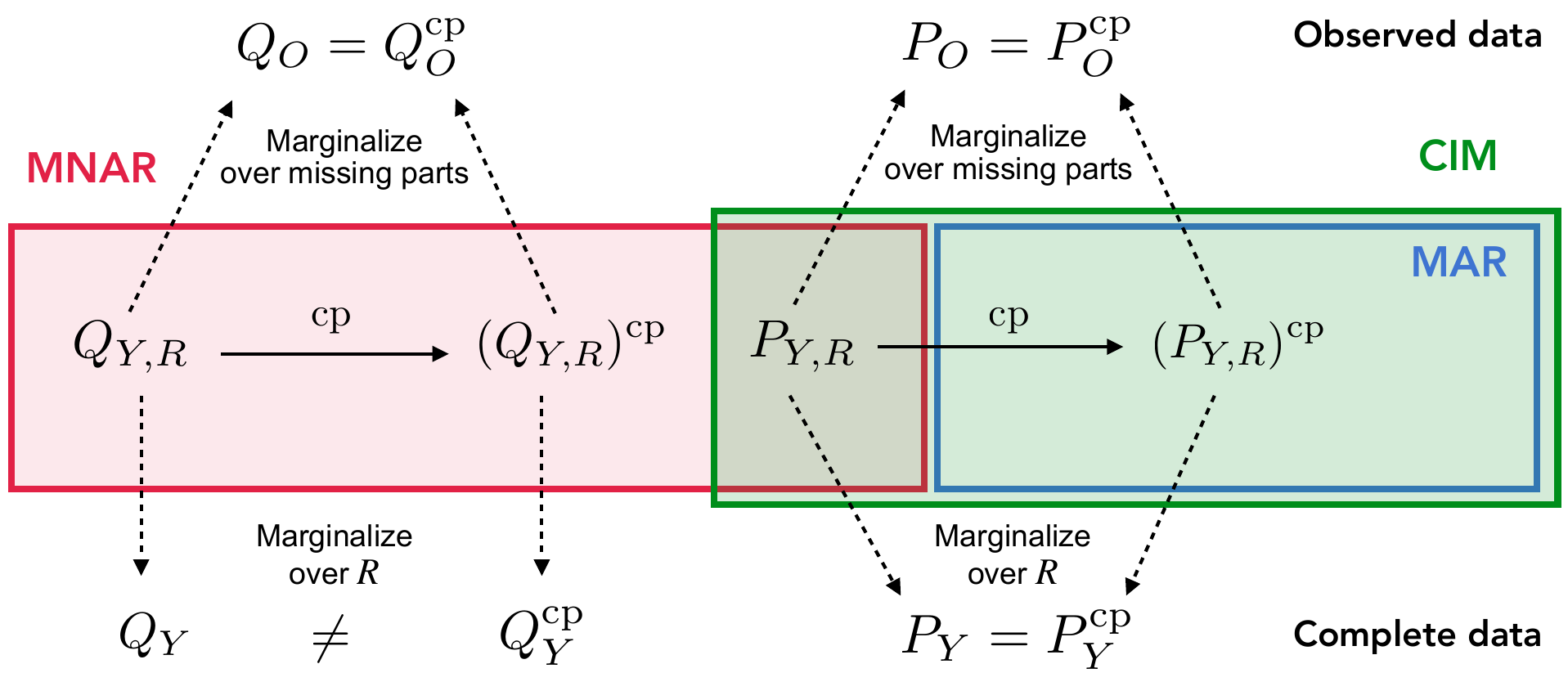}
\caption{Counterpart map and observed- and complete-data marginals.
The horizontal arrows labelled ``cp'' denote \eqref{eq:kano-counterpart-law}.
The upward arrows marginalize over the missing components while retaining
$R$, giving $Q_O=Q_O^{\mathrm{cp}}$ and $P_O=P_O^{\mathrm{cp}}$.
The downward arrows marginalize over $R$: $Q_Y\ne Q_Y^{\mathrm{cp}}$
for the non-CIM law, whereas $P_Y=P_Y^{\mathrm{cp}}$ under CIM.}
\label{fig:counterpart-schematic}
\end{figure}

Figure~\ref{fig:counterpart-schematic} illustrates the two marginalizations.

 For a target $\Psi(P_Y)$, a CIM law and its MAR counterpart thus have the same target and every observed-data statistic $T_n(O_1,\ldots,O_n)$ has the same finite-sample distribution under the two laws. For efficiency, assume identification on the common model:
\begin{equation}
P_O=P'_O\ \Longrightarrow\ \Psi(P_Y)=\Psi(P'_Y).
\label{eq:target-identification}
\end{equation}
Equivalently, there is an observed-data functional $\psi$ with
\begin{equation}
\Psi(P_Y)=\psi(P_O).
\label{eq:identified-functional}
\end{equation}
We use the observed-data tangent-space and canonical-gradient formulation of semiparametric efficiency \citep{BickelEtAl1993}.

\begin{corollary}[Efficiency invariance]\label{cor:efficiency-transfer}
Suppose that $\Psi(P_Y)$ is identified in the sense of
\eqref{eq:target-identification} on the common model in
\eqref{eq:parameter-model-equivalence}, with
$\Psi(P_Y)=\psi(P_O)$ as in \eqref{eq:identified-functional}, and
suppose that $\psi$ is pathwise differentiable at
the true observed-data law $P_{O,0}$. Then the efficient influence functions under the CIM and MAR
models coincide $P_{O,0}$-almost surely: $\phi_{\mathrm{eff}}^{\mathrm{CIM}}(O)=\phi_{\mathrm{eff}}^{\mathrm{MAR}}(O)$.
Consequently, the semiparametric efficiency bounds are identical: $V_{\mathrm{eff}}^{\mathrm{CIM}}=V_{\mathrm{eff}}^{\mathrm{MAR}}$.
\end{corollary}

The larger joint-law model introduces no new observed score directions for an identified marginal target, so a doubly robust or locally efficient MAR procedure is the same observed-data statistic with the same validity under CIM. Its identification, nuisance-model and regularity conditions must hold in the MAR counterpart: a pattern-probability model there describes $\pi_r(y^{(r)})$ rather than the actual probability $\bbP(R=r\mid Y=y)$, and the same qualification applies to outcome-regression nuisances. Under additional restrictions on the mechanism, the invariance persists only when the restricted model remains closed under the counterpart construction. Uniqueness with both marginals fixed is not identification of $P_Y$ from $P_O$ (Section~\ref{sec:identification}).

\begin{remark}[Relation to likelihood ignorability]\label{rem:lu-copas}
Theorem~\ref{thm:estimation-transfer} does not conflict with the equivalence of ignorability and MAR established by \citet{LuCopas2004}, who define likelihood ignorability within a specified
selection-model family by requiring each pattern's selection contribution
to the observed likelihood to be free of the substantive parameter
throughout the family. Under parameter distinctness and completeness of
the response family, this property is equivalent to MAR.
Theorem~\ref{thm:estimation-transfer} transfers the sampling properties of
observed-data procedures through an MAR counterpart with the same
observed-data law and complete-data marginal. This argument requires
neither a parameterization of the actual mechanism nor parameter
distinctness, and does not require the original selection-model likelihood
to be proportional to the direct likelihood.
\end{remark}

These transfer guarantees concern identified functionals of the complete-data marginal, not pattern-specific quantities (see the end of Section~\ref{sec:relation-mar}).

\section{Testing joint compatibility of CIM and a full model}\label{sec:tests}
Theorem~\ref{thm:estimation-transfer} suggests comparing the complete-data
marginal with its counterpart marginal. We realize the latter by
conditional completion and compare the completed and fitted model distributions.

\subsection{Pseudo-complete-data equivalence for testing}\label{sec:equivalence}

For a candidate $\theta\in\Theta$ whose conditional distributions are
defined at the observed values, retain each unit's observed components
and draw its missing components from
$f(y^{(-r)}\mid Y^{(r)};\theta)$ when $R=r$.
Write $\widetilde Y_\theta$ for the resulting pseudo-complete vector in
its original coordinate order,
$\widetilde P_{Y,\theta}=\mathcal L(\widetilde Y_\theta)$ for its
\emph{completed distribution}, and
$\widetilde P_{Y,R,\theta}=\mathcal L(\widetilde Y_\theta,R)$.
The observed-data law $P_O$ is held fixed as $\theta$ varies;
completion preserves the observed values and pattern, so its induced
observed-data law satisfies $\widetilde P_{O,\theta}=P_O$.
At the true complete-data parameter this construction realizes the
counterpart in Definition~\ref{def:counterpart}.

\begin{theorem}[Pseudo-complete-data characterization of CIM]
\label{thm:equivalence}
The joint law $\widetilde P_{Y,R,\theta_0}$ has density
\begin{align}
\widetilde p(y,r;\theta_0)
=\rho_r f_r(y^{(r)}\mid R=r) f(y^{(-r)}\mid y^{(r)};\theta_0)
=\pi_r(y^{(r)})f(y;\theta_0),
\quad r\in\mathcal R.
\label{eq:pseudo-counterpart-law}
\end{align}
Then, writing $\widetilde f(\cdot;\theta_0)$ for the density of $\widetilde Y_{\theta_0}$, equivalently the complete-data marginal density of $\widetilde P_{Y,R,\theta_0}$,
\begin{equation}
\widetilde P_{Y,R,\theta_0}=\mathcal L(\widetilde Y_{\theta_0},R)=(P_{Y,R})^{\mathrm{cp}},
\qquad
\widetilde f(y;\theta_0)=f(y;\theta_0)(1+F(y)).
\label{eq:pseudo-counterpart-identity}
\end{equation}
By \eqref{eq:pseudo-counterpart-identity} and
Theorem~\ref{thm:estimation-transfer}, CIM holds if and only if $\widetilde P_{Y,\theta_0}=P_{Y,\theta_0}$.
\end{theorem}

The null hypothesis is the joint statement
\begin{equation}
H_0:\ P_Y=P_{Y,\theta_0}
\quad\text{and}\quad
\widetilde P_{Y,\theta_0}=P_{Y,\theta_0}
\qquad\text{for some }\theta_0\in\Theta,
\label{eq:distributional-null}
\end{equation}
against the alternative under which no parameter satisfies both
equalities. The first equality specifies the full model; given it,
the second is CIM. Rejection indicates failure of either component
or both and cannot distinguish between them.
Every observed-data test sees only the observable counterpart of
\eqref{eq:distributional-null}, $H_0^{\mathrm{obs}}:\ \widetilde P_{Y,\theta}=P_{Y,\theta}$ for some $\theta\in\Theta$, with completion applied to $P_O$.
$H_0$ implies $H_0^{\mathrm{obs}}$, but a non-CIM law with an
observationally equivalent CIM explanation also satisfies
$H_0^{\mathrm{obs}}$ and cannot be distinguished from that explanation
by any observed-data test
(Supplementary Remark~\ref{rem:observational-equivalence-power}).
Theorem~\ref{prop:rkhs-limit} extends its size control to
$H_0^{\mathrm{obs}}$ whenever that CIM explanation satisfies its
conditions, and Section~\ref{sec:identification} gives conditions
excluding such observational equivalence.

The procedure uses the direct-likelihood estimate $\widehat\theta$
from \eqref{eq:observed-mle}. Under $H_0$, Supplementary Lemma~\ref{lem:consistency}
gives $\widehat\theta\overset{p}{\longrightarrow}\theta_0$ under its
conditions. Under the corresponding consistency conditions, its limit
under a fixed alternative is instead a population
maximizer $\theta^*\in\argmax_{\theta\in\Theta}\bar\ell(\theta)$,
where \eqref{eq:population-objective} defines $\bar\ell$.
Power therefore depends on the discrepancy at $\theta^*$.

\subsection{The fitted-kernel test on general complete-data spaces}\label{sec:continuous-test}

Use a bounded characteristic kernel $k:\cY\times\cY\to\R$ with reproducing kernel Hilbert space (RKHS)
$\cH$ to compare the distributions. This construction applies to
continuous, binary and ordinal data. Expectations without a parameter
subscript use $P_O$ and conditional completion; $\bbE_\theta$ uses
$P_{Y,\theta}$. Define $\mu_k(\theta)=\bbE_\theta(k(Y,\cdot))$, $m_{r,k}(u;\theta)=\bbE_\theta(k(Y,\cdot)\mid Y^{(r)}=u)$ and $g_{\cH}(\theta)=\bbE[m_{R,k}(Y^{(R)};\theta)]-\mu_k(\theta)$.
The conditional mean uses the realized pattern and requires only observed
coordinates. The discrepancy
$\Delta_{\cH}(\theta)=\|g_{\cH}(\theta)\|_{\cH}^2
=\MMD_k^2(\widetilde P_{Y,\theta},P_{Y,\theta})$,
the squared maximum mean discrepancy (MMD) between the completed and model distributions,
is zero exactly when the two distributions agree because $k$ is
characteristic; at $\theta_0$, this is equivalent to CIM.
For $O=(R,Y^{(R)})$ write
$\xi(O;\theta)=m_{R,k}(Y^{(R)};\theta)-\mu_k(\theta)$.
The fitted-kernel statistic is
\begin{align}
\widehat\xi_i
&=\xi(O_i;\widehat\theta),\qquad
\widehat g_{\cH}=\frac1n\sum_{i=1}^n\widehat\xi_i, \notag \\
T_{n,\cH}
&=n\|\widehat g_{\cH}\|_{\cH}^2
=\frac1n\sum_{i=1}^n\sum_{j=1}^n
\langle\widehat\xi_i,\widehat\xi_j\rangle_{\cH}.
\label{eq:rkhs-statistic}
\end{align}

Parameter estimation changes its first-order contribution to $\zeta(O)=\xi(O;\theta_0)+\dot g_\theta(\operatorname{IF}_\theta(O))$, where $\dot g_\theta=\partial g_{\cH}(\theta)/\partial\theta|_{\theta_0}$.
The derivative holds $P_O$ fixed, and $\operatorname{IF}_\theta$ is
the direct-likelihood estimator's influence function. Supplementary Assumption~\ref{ass:testing} states
the regularity conditions: i.i.d.\ units, a bounded kernel, asymptotic
linearity of the original and row-bootstrap fits, and smooth
conditional-completion maps with square-integrable derivative envelopes.
When the coordinates are standardized by estimated locations and
scales, this expansion uses their limiting values. Under the joint null,
the completed and model laws agree, so their kernel means agree for every
location and scale at fixed original-scale $\theta_0$, and the coordinate
derivative vanishes. Supplementary
Section~\ref{supp:estimated-scales} gives the regularity conditions
and the extension under alternatives.
Write $\widehat g_{\cH}^{(b)}$ for the embedding recomputed after whole-row
resampling and model refitting, as in Algorithm~\ref{alg:cim-calibration}.

\begin{theorem}[Fitted limit, bootstrap calibration and power]
\label{prop:rkhs-limit}
Suppose Supplementary Assumption~\ref{ass:testing} holds.
Under $H_0$ in \eqref{eq:distributional-null}, with $\theta_0$ the
parameter whose existence $H_0$ asserts, we have
\begin{equation}
\sqrt n\,\widehat g_{\cH}
=\frac1{\sqrt n}\sum_{i=1}^n\zeta(O_i)+o_p(1)
\ \overset{d}{\longrightarrow}\ \mathbb G_{\cH},
\label{eq:nuisance-expansion}
\end{equation}
where $\mathbb G_{\cH}$ is a centred Gaussian element with covariance
operator $\Var(\zeta(O))$. Consequently,
\begin{equation}
T_{n,\cH}\overset{d}{\longrightarrow}\|\mathbb G_{\cH}\|_{\cH}^2
=\sum_{j\geq1}\lambda_j Z_j^2,
\qquad Z_j\overset{\mathrm{i.i.d.}}{\sim}N(0,1),
\label{eq:weighted-chi-square}
\end{equation}
where $\lambda_j$ are the nonzero eigenvalues of this fitted covariance
operator. The conditional bootstrap law satisfies $d_{\mathrm{BL}}\{\mathcal L(\sqrt n(\widehat g_{\cH}^{(b)}-\widehat g_{\cH})\mid\cO_n),\mathcal L(\mathbb G_{\cH})\}\overset{p}{\longrightarrow}0$,
where $d_{\mathrm{BL}}$ is the bounded-Lipschitz distance between
probability laws on $\cH$, and the convergence in probability is with
respect to the original sample.
The centred refitted-bootstrap test is thus asymptotically level
$\alpha$ when the limiting quantile is nondegenerate and uniquely
determined at a continuity point.

Under a fixed alternative, assume the same regularity at its unique
pseudo-true parameter $\theta^*$. If
$\Delta_{\cH}(\theta^*)=\|g_{\cH}(\theta^*)\|_{\cH}^2>0$, then
$T_{n,\cH}/n\overset{p}{\longrightarrow}\Delta_{\cH}(\theta^*)$ and the rejection probability tends to one.
Finally, along an observed-data path differentiable in quadratic mean
through the null law, with score $s(O)$ and local parameter $n^{-1/2}$,
\begin{equation}
T_{n,\cH}\overset{d}{\longrightarrow}
\left\|\mathbb G_{\cH}+\bbE(\zeta(O)s(O))\right\|_{\cH}^2.
\label{eq:local-alternative}
\end{equation}
\end{theorem}

Concrete sufficient conditions for Gaussian full models are given
in Supplementary Corollary~\ref{cor:gaussian-regularity}.

\begin{algorithm}[htbp]
\caption{Centred refitted-bootstrap calibration of the fitted-kernel test}
\label{alg:cim-calibration}
\footnotesize
\textbf{Input:} Observed rows $O_1,\ldots,O_n$, a full-model family,
a fixed base kernel and declared coordinate rule, level $\alpha$, and
number of draws $B$.
\begin{enumerate}[1.]
\setlength{\itemsep}{3pt}
\item Fit $\widehat\theta$ by direct likelihood. When standardizing, compute
the means and sample SDs on the declared observed subsets and write
$\widehat\eta=(\widehat a,\widehat s)$. Set
$\widehat g_{\cH}=n^{-1}\sum_i\xi(O_i;\widehat\theta,\widehat\eta)$ and
$T_{n,\cH}=n\|\widehat g_{\cH}\|_{\cH}^2$; omit $\eta$ for fixed coordinates.
\item For each $b=1,\ldots,B$, independently resample $n$ whole observed rows.
Refit the model and recompute the declared means and SDs within that
sample to obtain $\widehat\theta^{(b)}$ and $\widehat\eta^{(b)}$.
\item Compute $\widehat g_{\cH}^{(b)}=n^{-1}\sum_i
\xi(O_i^{(b)};\widehat\theta^{(b)},\widehat\eta^{(b)})$ and
$T_b^{\mathrm{cal}}=n\|\widehat g_{\cH}^{(b)}-\widehat g_{\cH}\|_{\cH}^2$.
Both embeddings use the same base-kernel RKHS. In their cross inner
product, each uses its own sample's coordinate map
(Supplementary Section~\ref{supp:estimated-scales}).
\item Report
$\widehat p=(1+\sum_{b=1}^B\mathbb I\{T_b^{\mathrm{cal}}\ge
T_{n,\cH}\})/(B+1)$ and reject when $\widehat p\le\alpha$.
\end{enumerate}
\end{algorithm}

Influence-function-corrected Gaussian and Rademacher multiplier
calibrations, labelled MULT\_G and MULT\_R in the Supplementary Material, avoid repeated fitting; their
formulas and validity conditions are in Supplementary
Section~\ref{supp:multiplier}.

For an unpenalized interior Gaussian fit with free mean and covariance,
the completed averages of $Y$ and $YY^{\trans}$ equal
$\widehat\mu$ and
$\widehat\Sigma+\widehat\mu\widehat\mu^{\trans}$, even without CIM or
correct full-model specification; this is the Sundberg formula for the
observed-data score of an exponential family
\citep{Sundberg1974,DempsterLairdRubin1977}, restated as Supplementary
Proposition~\ref{prop:expfam-moment-matching}.
Thus fitting removes first- and second-moment discrepancies but need
not remove the characteristic-kernel discrepancy.
This moment agreement alone does not imply that the fitted mean and
covariance consistently estimate their complete-data counterparts.
For Gaussian full models with Gaussian kernels, the conditional and
marginal kernel means and their inner products have closed forms,
avoiding imputation Monte Carlo. Other conditional means can be
evaluated by fractional imputation \citep{Kim2011,SangKimLee2022}.
Supplementary Section~\ref{app:kernel-calculations} gives the Gaussian kernel calculations.

The test in \eqref{eq:rkhs-statistic} uses the marginal model embedding
as its reference. A version that conditions the reference embedding on
fully observed covariates is developed in Supplementary
Section~\ref{supp:centering}.

\section{Identification}\label{sec:identification}
The power result of Theorem~\ref{prop:rkhs-limit} requires a positive discrepancy after fitting. We now ask when a non-CIM law has no observationally equivalent CIM explanation within the specified full-model family.

\subsection{Identification from the observed-data law}\label{sec:cim-identification}
\begin{definition}[Identification of CIM]\label{def:cim-identification}
Whether a joint law satisfies CIM is identified from the observed-data
law within the full model $\mathcal M$ if, for every pair
$P_{Y,R},P'_{Y,R}$ under consideration with $P_Y,P'_Y\in\mathcal M$,
\begin{equation}
P_O=P'_O
\quad\Longrightarrow\quad
\left(
P_{Y,R}\text{ satisfies CIM}
\quad\Longleftrightarrow\quad
P'_{Y,R}\text{ satisfies CIM}
\right).
\label{eq:cim-membership-identification}
\end{equation}
The comparison includes laws that do not satisfy CIM.
When \eqref{eq:cim-membership-identification} holds, we say for brevity that CIM is identified.
\end{definition}

Identification of the complete-data marginal is sufficient for CIM to be identified, since equal observed and complete-data marginals determine the same $\pi_r$; identification of the entire joint law is unnecessary. A Gaussian full-model family alone does not suffice, as Supplementary Example~\ref{ex:gaussian-observational-equivalence} shows.

\subsection{Distributional sufficient conditions}\label{sec:distributional-identification}\label{sec:conditional-independence}
The following results exclude compatible CIM explanations in the stated family; restrictions on the true mechanism are not imposed on the comparison CIM law.

\begin{theorem}[Gaussian identification from complete-case tails]
\label{thm:gaussian-tail-identification}
Let $\mathcal M$ be the nondegenerate Gaussian distributions on
$\mathbb R^p$, and include the complete pattern $\mathbf1=(1,\ldots,1)$.
Suppose $P_Y\in\mathcal M$ and, for every $\epsilon>0$, there are
$c_\epsilon>0$ and $M_\epsilon<\infty$ such that
\begin{equation}
\pi_{\mathbf1}(y)=\bbP(R=\mathbf1\mid Y=y)
\geq c_\epsilon\exp(-\epsilon\|y\|^2)
\quad\text{for Lebesgue-a.e. }\|y\|\geq M_\epsilon.
\label{eq:gaussian-tail-main}
\end{equation}
If a CIM law $P'_{Y,R}$ with $P'_Y\in\mathcal M$ satisfies $P'_O=P_O$,
then $P'_Y=P_Y$ and $P_{Y,R}$ satisfies CIM. In particular, a non-CIM
law satisfying \eqref{eq:gaussian-tail-main} has no Gaussian--CIM
representation of its observed distribution. The remaining observation
patterns are arbitrary.
\end{theorem}

A uniformly positive complete-case probability, or affine multinomial-logistic selection, satisfies \eqref{eq:gaussian-tail-main}. This is a sufficient condition on an unobserved selection probability and cannot generally be checked from $P_O$. It is not necessary: normal--probit identification can follow from a different argument \citep{MiaoDingGeng2016}.

A second route uses variables observed in only one pattern and does not require complete cases. A nonempty coordinate block is \emph{exclusive} to $r$ if it is observed under $r$ and unobserved under every other pattern.

\begin{proposition}[Exclusive-variable property]
\label{prop:exclusive-variable}
Fix $r\in\mathcal R$ with $\rho_r>0$ and write $Y^{(r)}=(X,V)$, where $V$
is a nonempty block of coordinates exclusive to pattern $r$; the remaining
observed coordinates $X$ need not be observed outside pattern $r$. When
$r$ is incomplete, assume that the complete-data density is positive on a
product support. Under CIM, $\pi_r(X,V)=\bbP(R=r\mid X)$, or equivalently
\mbox{$V\perp\mathbb I\{R=r\}\mid X$}, and consequently
\begin{equation}
\mathcal L(V\mid X,R=r)=\mathcal L(V\mid X)
\label{eq:exclusive-block-main}
\end{equation}
for almost every $x$ with $\bbP(R=r\mid X=x)>0$.
\end{proposition}

This property concerns the event $R=r$, not independence of $V$ and the entire pattern $R$ given $X$, and does not imply MAR. Examples are in Supplementary Section~\ref{supp:exclusive-variable-proof-examples}.

\begin{theorem}[Identification by the shape of conditional observation probabilities]
\label{thm:conditional-shape-identification}
Use the exclusive-variable structure $(r,X,V)$ of
Proposition~\ref{prop:exclusive-variable}, with $V$ scalar. Suppose the
complete-data laws have positive densities on a common product support
and $P_Y\in\mathcal M$. Assume either that all induced conditional laws
$V\mid X=x$ are nondegenerate Gaussian, or that all are Student $t$
with arbitrary positive finite degrees of freedom. If, on a set of
$x$ values of positive probability, $\pi_r(x,\cdot)$ is positive,
not almost everywhere constant, and monotone, then no CIM law with marginal in
$\mathcal M$ has the same observed-data distribution.

\end{theorem}

Multivariate $t$ conditionals remain Student $t$ \citep{Ding2016}, allowing unknown location, scatter and finite degrees of freedom; the marginal law of $X$ may remain unspecified. Supplementary Section~\ref{supp:conditional-identification} gives extensions and an example without complete cases.

Supplementary Theorems~\ref{thm:logit-patterns}, \ref{thm:fixed-template-mixture-identification} and~\ref{thm:mixture-probit-separation} give further identification results for Student $t$ mixtures and Gaussian mixtures under their respective observation and component restrictions.

\section{Numerical studies}\label{sec:simulation}

Both experiments use the statistic in \eqref{eq:rkhs-statistic} and
the centred refitted bootstrap in Algorithm~\ref{alg:cim-calibration},
with $B=499$ draws per simulated dataset. Gaussian kernels are bounded
and characteristic, so the discrepancy vanishes only when the completed
and model laws agree, and under a Gaussian full model their conditional
and marginal kernel means have closed forms (Section~\ref{sec:tests}).
A single bandwidth is sensitive mainly to departures at its own scale,
and a positive combination of characteristic kernels is again
characteristic. To assess discrepancies at several scales without
tuning a bandwidth, we therefore average Gaussian kernels at three fixed
bandwidths:
$k(y,y')=\frac13\sum_{\sigma\in\{1,2,4\}}
\exp\{-\|y-y'\|^2/(2\sigma^2)\}$, with bandwidths in population
standard-deviation units. The generating mechanisms are given in
Supplementary Section~\ref{supp:moved-simulation-design}.

\subsection{Experiment A: invariant inference and null calibration}
\label{sec:simulation-transfer}
Generate $Y=(Y_1,Y_2,Y_3,Y_4)\sim N_4(0,\Sigma)$, with
$\Sigma_{jk}=0.5^{|j-k|}$ and always-observed $C=Y_1$. The patterns
are $1111$, $1100$, $1010$ and $1000$. A0 assigns each probability
$1/4$; A1 uses MAR probabilities depending on $C$. A2 redistributes
probability between $1100$ and $1010$ by opposite bounded terms
$b(C)\{2\expit(2U)-1\}$, where $U=(Y_4-0.5Y_3)/\sqrt{0.75}$ is independent of
$(C,Y_2,Y_3)$. These terms integrate to zero, giving A1 and A2 the
same observed law and complete-data marginal although A2 is MNAR.
Every fixed observed-data procedure therefore has the same sampling
distribution in A1 and A2, illustrating
Theorem~\ref{thm:estimation-transfer}.

For each A0--A2 law and each sample size $n=500,1000,2000$,
we run 1,000 independent Monte Carlo iterations.
The Gaussian fit estimates all four means and ten covariance parameters
without imposing the generating AR(1) structure. Targets are
$\bbE(Y_4)=0$ and $\Cov(Y_2,Y_3)=0.5$; their percentile intervals use
the same bootstrap refits as the test.

The mean and covariance estimates have small biases and decreasing root
mean squared errors (RMSEs) under all three laws (Table~\ref{tab:A-main}).
Across the nine null settings,
rejection rates range from $3.6\%$ to $6.7\%$ at the nominal $5\%$ level.
At $n=2000$, the rates are $6.7\%$ under A1 and $4.9\%$ under A2;
the pooled rate is $5.8\%$ for the two independent sets of replicates
from the same observed-data law.

\begin{table}[t]
\centering
\tablebodysetup
\setlength{\tabcolsep}{2.4pt}
\caption{Experiment A: joint-null rejection and target inference.
Each row summarizes 1,000 independent Monte Carlo iterations. Rejection
of the joint null at the $5\%$ level and coverage of the $95\%$
percentile intervals are percentages; parentheses give Monte Carlo
standard errors in percentage points. The test and intervals use the
same $B=499$ whole-row bootstrap refits.}
\label{tab:A-main}
\begin{tabular*}{\textwidth}{@{\extracolsep{\fill}}llrrrrr@{}}
\toprule
& & & \multicolumn{2}{c}{Mean $Y_4$} & \multicolumn{2}{c}{$\Cov(Y_2,Y_3)$} \\
\cmidrule(lr){4-5}\cmidrule(l){6-7}
Law & $n$ & Rejection & Bias / RMSE & Coverage & Bias / RMSE & Coverage \\
\midrule
A0 & 500 & 4.3 (0.6) & 0.001 / 0.085 & 94.0 (0.8) & 0.002 / 0.078 & 94.7 (0.7) \\
A0 & 1000 & 4.6 (0.7) & -0.002 / 0.060 & 93.9 (0.8) & -0.003 / 0.054 & 95.3 (0.7) \\
A0 & 2000 & 4.5 (0.7) & 0.000 / 0.044 & 93.1 (0.8) & 0.000 / 0.039 & 94.6 (0.7) \\
\addlinespace[2pt]
A1 & 500 & 4.2 (0.6) & 0.002 / 0.086 & 93.9 (0.8) & 0.000 / 0.079 & 94.7 (0.7) \\
A1 & 1000 & 4.8 (0.7) & -0.001 / 0.062 & 94.7 (0.7) & 0.002 / 0.059 & 93.6 (0.8) \\
A1 & 2000 & 6.7 (0.8) & 0.001 / 0.042 & 94.4 (0.7) & -0.002 / 0.040 & 94.9 (0.7) \\
\addlinespace[2pt]
A2 & 500 & 3.6 (0.6) & 0.000 / 0.089 & 93.3 (0.8) & 0.002 / 0.081 & 93.9 (0.8) \\
A2 & 1000 & 4.6 (0.7) & 0.001 / 0.061 & 93.5 (0.8) & -0.001 / 0.056 & 93.7 (0.8) \\
A2 & 2000 & 4.9 (0.7) & 0.002 / 0.042 & 95.4 (0.7) & 0.002 / 0.040 & 93.6 (0.8) \\
\bottomrule
\end{tabular*}
\end{table}

\subsection{Experiment B: departures from CIM and from the full model}
\label{sec:simulation-power}
We use 500 independent simulations per setting at each of the three
sample sizes. B1 and B2 retain the Gaussian complete-data law and
patterns of Experiment~A. Both exponentially tilt the MAR pattern
probabilities: B1 uses bounded functions of unobserved coordinates
and B2 uses bounded functions of their squares, with strengths
$\gamma=0.5,1$. Both depend on $Y_4$ and violate the exclusive-variable
identity~\eqref{eq:exclusive-block-main}. Their complete-case probability
is at least $0.05e^{-\gamma}$, so
Theorem~\ref{thm:gaussian-tail-identification} excludes every
Gaussian--CIM explanation. B1 also satisfies
Theorem~\ref{thm:conditional-shape-identification}'s monotonicity
condition; B2 does not. Supplementary Corollary~\ref{prop:gaussian-separated-alternatives} shows that the resulting discrepancy remains positive after fitting.

Two other mechanisms assign each pattern probability $1/4$
independently of $Y$, so CIM holds. Here the departure is instead
from the Gaussian full model: a Gaussian mixture and a $t_{10}$ law
each have the same mean and covariance as the Gaussian law in
Experiment~A.

Table~\ref{tab:B-main} distinguishes the two sources of departure.
Under non-CIM selection, rejection rates increase with sample size and
tilt strength in both B1 and B2. Under MCAR, the test detects the
heavy-tailed law with increasing frequency, whereas rejection rates for
the moment-matched mixture stay close to the nominal level over the
sample sizes examined. Thus rejection can reflect the full model alone;
nonrejection does not establish that the Gaussian model and CIM both hold.
Supplementary Table~\ref{tab:B-inference} reports bias, RMSE
and 95\% interval coverage for both targets across all 18 settings.

\begin{table}[t]
\centering
\tablebodysetup
\setlength{\tabcolsep}{2.4pt}
\caption{Experiment B: rejection of joint Gaussian--CIM compatibility
at the $5\%$ level. Entries are rejection percentages across 500
independent Monte Carlo iterations, with Monte Carlo standard errors
in percentage points in parentheses. Each test uses $B=499$ centred
whole-row bootstrap refits.}
\label{tab:B-main}
\begin{tabular*}{\textwidth}{@{\extracolsep{\fill}}lrrr@{}}
\toprule
Law & $n=500$ & $n=1000$ & $n=2000$ \\
\midrule
\multicolumn{4}{@{}l}{\textit{Gaussian complete-data law, non-CIM selection}} \\
B1, $\gamma=0.5$ & 5.2 (1.0) & 6.8 (1.1) & 8.2 (1.2) \\
B1, $\gamma=1$ & 6.6 (1.1) & 10.4 (1.4) & 19.8 (1.8) \\
B2, $\gamma=0.5$ & 6.2 (1.1) & 8.0 (1.2) & 12.6 (1.5) \\
B2, $\gamma=1$ & 11.2 (1.4) & 20.8 (1.8) & 42.4 (2.2) \\
\addlinespace
\multicolumn{4}{@{}l}{\textit{Non-Gaussian complete-data law, MCAR selection}} \\
Mixture & 4.2 (0.9) & 4.2 (0.9) & 4.8 (1.0) \\
$t_{10}$ & 62.0 (2.2) & 97.6 (0.7) & 100.0 (0.0) \\
\bottomrule
\end{tabular*}
\end{table}

\section{Real-data applications}\label{sec:real-data}

\subsection{SUPPORT2 laboratory measurements}\label{sec:support2-application}
The SUPPORT study investigated seriously ill hospitalized adults
\citep{KnausEtAl1995}. We use the public SUPPORT2 ARF/MOSF with Sepsis
group of $3515$ patients \citep{HarrellSupportData}, first examining
study-day-3 serum albumin (\texttt{alb}) and oxygenation
($\mathrm{PaO}_2/\mathrm{FiO}_2$, \texttt{pafi}). Pattern counts are
$11:1962$, $10:293$, $01:1044$ and $00:216$; all records remain in
the fit and resampling. Observed mean albumin is $3.153$ in pattern
$10$ versus $2.722$ in pattern $11$. Blood-gas measurement assesses
oxygenation and cardiopulmonary severity \citep{DavisEtAl2013}, so
recording related to incompletely captured respiratory status is a
plausible explanation for incompatibility.

Let $Y=(\sqrt{\mathrm{alb}},\sqrt{\mathrm{pafi}})$. We analyse
$(Y_j-\widehat a_j)/\widehat s_j$, where the means $(1.6348,14.527)$
and SDs $(0.2218,3.507)$ use the $1962$ paired observations.
We fit a bivariate Gaussian working model with all five parameters free
and use the equally weighted Gaussian-kernel mixture from Section~\ref{sec:simulation},
with bandwidths $1,2,4$. Each bootstrap sample re-estimates the paired
means and SDs. Thus changes in the origin or units of a transformed
laboratory measurement leave both the observed statistic and its
bootstrap calibration unchanged (Supplementary
Section~\ref{supp:estimated-scales}).

\subsection{Estimation, joint-compatibility testing and additional diagnostics}
\label{sec:support2-bivariate-test}
With $B=1999$ draws, the refitted bootstrap rejects joint
Gaussian--CIM compatibility: $T=1.179$ and the 95\% critical value is
$0.1405$. The Gaussian and Rademacher multiplier critical values are
$0.1432$ and $0.1428$, respectively. All three calibrations give
$\widehat p=0.0005$, the smallest attainable value with $1999$ draws
(Supplementary Table~\ref{tab:support2-tests}). Supplementary
Section~\ref{supp:support2-targets} reports descriptive summaries
of the original-scale laboratory means.

For conditional sensitivity analyses, we also condition on age, acute physiology score, coma score and mean
blood pressure, using the $3513$ patients with these covariates observed.
We use the released physiology score, including its imputed components. Gaussian conditional models use four laboratories
(albumin, oxygenation, creatinine and blood urea nitrogen), add
bilirubin and white blood cell count in a six-laboratory specification,
or replace the four-laboratory linear mean with a fixed-knot cubic mean.
The covariate marginal is unspecified; all laboratory patterns remain
in the analysis. Conditional kernel means remove each individual's
model-implied reference embedding. Model and scale details are in
Supplementary Section~\ref{sec:support2-conditional}.
All three specifications reject their respective joint nulls
($\widehat p=0.0005$ for each; Supplementary
Table~\ref{tab:support2-tests}).

A descriptive density reconstruction locates the bivariate discrepancy
(Supplementary Section~\ref{sec:support2-density}). Joint rejection
does not distinguish full-model misspecification from failure of CIM.

\subsection{Childhood growth in the Terneuzen Birth Cohort}
In an exploratory analysis of the public Terneuzen Birth Cohort subset,
we estimated the mean change in body mass index standard deviation score
between visits near ages 2 and 6, a growth period studied by
\citet{DeKroon2010}. Of 306 children, 91 had both scores, 61 had only
the earlier score and 27 only the later score. The mean change was estimated as $-0.475$
(bootstrap SE $0.111$) from the 91 complete pairs. Including the 88
children with one score in the Gaussian direct likelihood gave
$-0.489$ (SE $0.100$). The diagnostic did not reject joint
Gaussian--CIM compatibility ($p=0.2355$; Supplementary
Section~\ref{sec:tbc}).

\section{Concluding Remarks}\label{sec:conclusion}
CIM relaxes the patternwise MAR restrictions to the single normalization
identity $\sum_{r\in\mathcal R}\pi_r(y^{(r)})=1$, which is equivalent to
invariance of the complete-data distribution under conditional completion
with the true full model. This shows that some departures from MAR can be
immaterial for inference on identified features of the complete-data
distribution.

The proposed diagnostic assesses the complete-data model and CIM jointly. It can therefore reject because the
full model is misspecified even when CIM holds. This matters when a
distributional assumption is used as a working approximation, as a Gaussian
model may be in factor analysis and structural equation modeling.
Interest in these settings may lie primarily in mean and covariance
structures, and a rejection leaves unresolved whether ignoring missingness
compromises inference for these targets. An important direction is
therefore to assess when ignoring missingness remains appropriate for
specified targets under a working model. Developing such an approach
requires balancing robustness to distributional misspecification with
identification: enlarging the full-model family may improve approximation,
but can also admit observationally equivalent CIM and non-CIM explanations.

Further work should develop substantively plausible parametric
missingness models whose induced pattern-conditional probabilities satisfy
the CIM normalization. The low-dimensional constructions in this paper
demonstrate feasibility, but the normalization couples probabilities
conditioned on different sets of observed variables and does not follow
from ordinary normalization of a missingness model. Extending these
constructions to richer pattern structures while retaining interpretable
parameters and valid probabilities would help connect CIM to plausible
observation processes in applications.

\section*{Author contributions statement}
K.M. contributed to all aspects of the study, including conceptualization,
theory, methodology, implementation, analyses, and writing. Y.K. contributed
to conceptualization, the formulation of research questions centred on CIM,
and research direction. S.S. contributed to developing the estimation and
testing methods. 

\section*{Data availability statement}
The SUPPORT2 data analysed in Section~\ref{sec:real-data} are publicly
available from Vanderbilt Biostatistics \citep{HarrellSupportData}.
The public Terneuzen Birth Cohort subset is available as \texttt{tbc} in
the R package \texttt{mice} (\url{https://amices.org/mice/reference/tbc.html}).
The code and numerical materials reproducing all results in
Sections~\ref{sec:simulation} and~\ref{sec:real-data} and the Supplementary
Material are archived as version 1.0.1 in Zenodo
\citep{MorikawaEtAl2026Reproducibility} at
\url{https://doi.org/10.5281/zenodo.24096545}.
The archive includes acquisition instructions and checksum-verified
retrieval scripts for both datasets; third-party individual-level data
are obtained from their original providers rather than redistributed.

\section*{Acknowledgements}
ChatGPT (OpenAI) assisted with language editing, code development, numerical analyses, and searches for literature and public datasets. It also suggested several candidate names for the condition in Definition 1 (Completion-invariant missingness). Claude (Anthropic) assisted with language editing. The authors are responsible for the content of the manuscript.

The authors also thank Professor Jae Kwang Kim of Iowa State University for his many helpful comments, both before and during the preparation of this manuscript.

\section*{Funding}
The authors were partially supported by JSPS KAKENHI Grant Numbers JP24K20743 and JP23H00466 (K.M.), JP25K15032 (Y.K.), and JP25H00546 and JP24K21420 (S.S.).

\bibliographystyle{abbrvnat}
\bibliography{refs}

\clearpage
\phantomsection
\addcontentsline{toc}{section}{Supplementary Material}
\begin{center}
{\LARGE Supplementary Material}
\end{center}
\bigskip

\suppcounter{section}
\suppcounter{equation}
\suppcounter{table}
\suppcounter{figure}
\suppcounter{algorithm}
\suppcounter{theorem}
\suppcounter{proposition}
\suppcounter{lemma}
\suppcounter{corollary}
\suppcounter{definition}
\suppcounter{assumption}
\suppcounter{remark}
\suppcounter{example}

This supplement accompanies the main paper and uses its notation;
``full model'' denotes a model for the complete-data distribution.
Because \citet{Kano2014} is written in Japanese, every result taken from that paper is restated here with a complete proof.
Section~\ref{supp:proofs} collects the regularity conditions and the proofs of the main-paper results.
Section~\ref{supp:testing-details} gives the calibration and computational details of the fitted-kernel test.
Section~\ref{supp:identification-separation} gives further identification results and examples.
Section~\ref{supp:revised-protocol} gives simulation details, calibration checks and target-inference summaries.
Sections~\ref{supp:support2-checks} and~\ref{sec:tbc} describe the SUPPORT2 and Terneuzen Birth Cohort applications.

\section{Theoretical results and proofs}\label{supp:proofs}\label{supp:framework-inference-details}\label{app:core-proofs}

\subsection{Two-pattern equivalence}
\label{supp:two-patterns-gaussian-example}

The distinction between CIM and MAR is already visible within monotone missingness. 
If $\mathcal R=\{\mathbf1,r\}$ with $r\ne\mathbf1$, CIM gives
\[
\bbP(R=r\mid Y=y)
=1-\pi_{\mathbf1}(y)
=\pi_r(y^{(r)})
\qquad P_{Y,\theta_0}\text{-a.s.},
\]
which is the only nontrivial MAR restriction. Together with
MAR $\Rightarrow$ CIM, this proves equivalence in this case, as shown by
\citet[Proposition~4.2]{Kano2014}.

The complete-pattern condition is essential; having two patterns alone
does not imply equivalence. Proposition~\ref{prop:three-pattern-counterexample} shows that even
monotone missingness no longer gives equivalence once a second incomplete
pattern is present.

\subsection{Direct-likelihood consistency}
\label{supp:direct-likelihood-consistency}

\begin{lemma}[Consistency of direct-likelihood estimation under CIM; \citeauthor{Kano2014},
\citeyear{Kano2014}]\label{lem:consistency}
Suppose that $f(\cdot;\theta_0)$ is correctly specified and CIM holds.
Assume that $\bar\ell(\theta_0)$ is finite, the log likelihood ratios below are
integrable, $P_{Y,\theta}\ll P_{Y,\theta_0}$ for every
$\theta\in\Theta$, and
\[
\sup_{\theta\in\Theta}
\left|n^{-1}\ell_n^{\mathrm{obs}}(\theta)-\bar\ell(\theta)\right|=o_p(1).
\]
Suppose also that the population maximizer is well separated: for every
$\delta>0$,
\[
\sup_{\theta\in\Theta:\|\theta-\theta_0\|\geq\delta}\bar\ell(\theta)
<\bar\ell(\theta_0).
\]
Then any maximizer in \eqref{eq:observed-mle} satisfies $\widehat\theta\overset{p}{\longrightarrow}\theta_0$.
\end{lemma}

CIM enters only through the population criterion: it makes $\theta_0$
a maximizer of $\bar\ell$, and the remaining conditions of
Lemma~\ref{lem:consistency} convert population maximization into
consistency. The proof is in Section~\ref{app:theoretical-proofs}.

\subsection{The exclusive-variable property}
\label{supp:exclusive-variable-proof-examples}
\begin{proof}[Proof of Proposition~\ref{prop:exclusive-variable}]
Under CIM,
\[
\pi_r(X,V)=1-\sum_{s\in\mathcal R\setminus\{r\}}\pi_s(Y^{(s)})
\qquad P_Y\text{-a.s.}
\]
Because $V$ is exclusive to pattern $r$, the right-hand side is a
function of the coordinates outside $V$. If $r$ is the complete pattern,
those coordinates are exactly $X$, so $\pi_r(X,V)$ is a function of $X$
alone. If $r$ is incomplete, the right-hand side may also involve
coordinates unobserved under $r$, whereas the left-hand side depends only
on $(X,V)$; positivity of the complete-data density on a product support
and Fubini's theorem then imply that both sides are functions of their
common coordinates $X$ alone. In either case
$\pi_r(X,V)=\bbP(R=r\mid X)$, that is, $V\perp I\{R=r\}\mid X$, and
Bayes' formula gives
\[
f(v\mid x,R=r)
=\frac{\pi_r(x,v)f(v\mid x)}{\bbP(R=r\mid X=x)}
=f(v\mid x)
\]
wherever $\bbP(R=r\mid X=x)>0$, which is
\eqref{eq:exclusive-block-main}.
\end{proof}

For the two-pattern setting $\mathcal R=\{11,10\}$, take $X=Y_1$, $V=Y_2$, and $r=11$. 
Since CIM and MAR coincide in this case, \eqref{eq:exclusive-block-main} recovers the familiar identity $\mathcal L(Y_2\mid Y_1,R=11)=\mathcal L(Y_2\mid Y_1)$.
The same reasoning applies to the nonmonotone patterns $1111,1100,1010,1000$: with $V=Y_4$, $X=(Y_1,Y_2,Y_3)$, and $r=1111$, \eqref{eq:exclusive-block-main} still holds under CIM, even though the other pattern probabilities may depend on unobserved values. 
Thus CIM can retain this particular conditional-independence interpretation while allowing MNAR mechanisms.

\subsection{Proof of Theorem~\ref{thm:estimation-transfer}}\label{app:equivalence-proof}
\begin{proof}[Proof of Theorem~\ref{thm:estimation-transfer}]
We first verify that the counterpart is a probability law preserving
the observed-data distribution. We then prove (i)$\Longleftrightarrow$(iii)
and (i)$\Longleftrightarrow$(ii), followed by uniqueness of the MAR
counterpart and equality of the attainable pairs $(P_O,P_Y)$.

\emph{Validity and preservation of the observed-data law.}
The counterpart in Definition~\ref{def:counterpart} is a probability
law because
\[
\sum_{r\in\mathcal R}
\int \pi_r(y^{(r)})\,P_Y(dy)
=
\sum_{r\in\mathcal R} \bbP(R=r)
=1.
\]
For every measurable set $A$ in the range of $Y^{(r)}$,
\begin{align*}
(P_{Y,R})^{\mathrm{cp}}(R=r,Y^{(r)}\in A)
&=\int_A \pi_r(u)\,P_{Y^{(r)}}(du)\\
&=\bbP(R=r,Y^{(r)}\in A),
\end{align*}
so $P_O^{\mathrm{cp}}=P_O$.

\emph{Equivalence of (i) and (iii): CIM and preservation of $P_Y$.}
Summing \textup{(\ref{eq:kano-counterpart-law})} over $r$ gives
\[
P_Y^{\mathrm{cp}}(dy)
=\Bigl\{\sum_{r\in\mathcal R}\pi_r(y^{(r)})\Bigr\}P_Y(dy)
=\{1+F(y)\}P_Y(dy),
\]
with $F$ as in \textup{(\ref{eq:F})}. Hence
$P_Y^{\mathrm{cp}}=P_Y$ if and only if $F=0$ $P_Y$-almost surely, which is
exactly CIM. This proves the equivalence of (i) and (iii).

\emph{Implication (i)$\Rightarrow$(ii): CIM makes the counterpart MAR.}
Under CIM, the previous display shows $P_Y^{\mathrm{cp}}=P_Y$.
Dividing the joint counterpart law by this marginal gives
\[
(P_{Y,R})^{\mathrm{cp}}(R=r\mid Y=y)
=\pi_r(y^{(r)}),
\]
so the counterpart mechanism is MAR. Thus (i) implies (ii).

\emph{Implication (ii)$\Rightarrow$(i): an MAR counterpart implies CIM.}
Suppose that $(P_{Y,R})^{\mathrm{cp}}$ is MAR. We will show
$\bbE[F(Y)^2]=0$, which forces $F=0$ almost surely.
First, on the event
$\pi_r(Y^{(r)})=0$, the nonnegative conditional probability
$\bbP(R=r\mid Y)$ also vanishes $P_Y$-almost surely, by
iterated conditioning: its conditional expectation given $Y^{(r)}$
is $\pi_r(Y^{(r)})$, and a nonnegative variable with
conditional mean zero must vanish there. Hence the difference
$F_r(Y)=\pi_r(Y^{(r)})
-\bbP(R=r\mid Y)$ is zero there. If
$1+F(Y)=\sum_{r\in\mathcal R}\pi_r(Y^{(r)})$ were zero on an event of
positive probability, all full-data pattern probabilities would vanish
there, contradicting their sum being one. Thus $1+F(Y)>0$
$P_Y$-almost surely, and $P_Y^{\mathrm{cp}}$ and $P_Y$ have the same
null sets.

Dividing the joint counterpart law by
$P_Y^{\mathrm{cp}}(dy)=\{1+F(y)\}P_Y(dy)$ gives its conditional pattern
probability. Since the counterpart is MAR, this probability is a
function of $Y^{(r)}$ alone:
\[
(P_{Y,R})^{\mathrm{cp}}(R=r\mid Y^{(r)})
=\frac{\pi_r(Y^{(r)})}{1+F(Y)}
\qquad P_Y\text{-almost surely}.
\]
On the event $\pi_r(Y^{(r)})>0$, the denominator can be
recovered from the ratio of the two conditional probabilities, so
$F(Y)$ is a function of $Y^{(r)}$ there. Consequently,
$\mathbb I(\pi_r(Y^{(r)})>0)F(Y)$ is a bounded
$\sigma(Y^{(r)})$-measurable function. Since $F_r=0$ on the
complementary event $\{\pi_r(Y^{(r)})=0\}$, inserting
the indicator does not change $F(Y)F_r(Y)$. We can therefore condition
on $Y^{(r)}$ and use
$\bbE(F_r(Y)\mid Y^{(r)})=0$,
\[
\bbE(F(Y)F_r(Y))
=\bbE\!\left[
\mathbb I(\pi_r(Y^{(r)})>0)F(Y)
\bbE(F_r(Y)\mid Y^{(r)})\right]=0.
\]
All terms are integrable because there are finitely many patterns.
Since $F=\sum_{r\in\mathcal R} F_r$, summing over $r$ yields
\[
\bbE(F(Y)^2)
=\sum_{r\in\mathcal R}\bbE(F(Y)F_r(Y))=0.
\]
Thus CIM holds, proving (ii)$\Rightarrow$(i) and completing the
equivalence of (i), (ii), and (iii).

\emph{Existence and uniqueness with both marginals fixed.}
Suppose first that an MAR law $P'_{Y,R}$ satisfies
$P'_O=P_O$ and $P'_Y=P_Y$. The first equality fixes the
observed-pattern subdistributions, and the second fixes the marginal
distribution of every observed subvector. Their ratio therefore gives
the same $\pi_r$ under both laws on the corresponding marginal support.
Since $P'_{Y,R}$ is MAR,
\[
P'_{Y,R}(R=r\mid Y=y)
=\pi_r(y^{(r)}),
\]
so $P'_{Y,R}=(P_{Y,R})^{\mathrm{cp}}$. Hence the counterpart is MAR,
which gives (ii), and therefore (i)--(iii).

Conversely, when (i)--(iii) hold, the counterpart is an MAR law with
the required observed-data and complete-data marginals. The preceding
argument also proves uniqueness.

\emph{Equality of the attainable pairs $(P_O,P_Y)$.}
The two sides of \textup{(\ref{eq:parameter-model-equivalence})}
are sets of marginal pairs, so we prove the two set inclusions.
Every MAR law satisfies CIM. Indeed, under the MAR condition $\bbP(R=r\mid Y=y)=\pi_r(y^{(r)})$,
\[
\sum_{r\in\mathcal R}\pi_r(y^{(r)})
=
\sum_{r\in\mathcal R}
\bbP(R=r\mid Y=y)
=1
\qquad P_Y\text{-a.s.}
\]
Thus the right-hand side of
\textup{(\ref{eq:parameter-model-equivalence})} is contained in the left-hand
side.

For the reverse inclusion, fix
$P_{Y,R}\in\mathcal P_{\mathrm{CIM}}(\mathcal M)$ and let
$(P_{Y,R})^{\mathrm{cp}}$ be the law in
\textup{(\ref{eq:kano-counterpart-law})}. The preceding argument gives
\[
(P_{Y,R})^{\mathrm{cp}}\in\mathcal P_{\mathrm{MAR}}(\mathcal M),
\qquad
P_O^{\mathrm{cp}}=P_O,
\qquad
P_Y^{\mathrm{cp}}=P_Y.
\]
Consequently,
\[
\bigl(P_O^{\mathrm{cp}},
P_Y^{\mathrm{cp}}\bigr)
=
\bigl(P_O,P_Y\bigr),
\]
which proves the reverse inclusion.
This last step uses existence of the MAR counterpart preserving both
marginals; uniqueness is the separate assertion established above.
\end{proof}

\subsection{Proof of Corollary~\ref{cor:efficiency-transfer}}\label{app:efficiency-proof}
\begin{proof}[Proof of Corollary~\ref{cor:efficiency-transfer}]
Projecting Theorem~\ref{thm:estimation-transfer} onto its first
coordinate shows that the CIM and MAR models induce the same set of
observed-data laws. The theorem also preserves the complete-data
marginal. The identification assumption in
\textup{(\ref{eq:target-identification})} and its representation in
\textup{(\ref{eq:identified-functional})} therefore define the same parameter map
$\psi$ on this common model.

Consider any regular one-dimensional observed-data submodel $P_{O,t}$ through
the true law $P_{O,0}$, with score $s\in L_0^2(P_{O,0})$.
Expectations in this proof are under $P_{O,0}$. Because the model sets
coincide, the admissible observed-data submodels, and hence the closed
linear span of their scores, are identical under CIM and MAR. Denote
this common tangent space by $\mathcal T$. Along each such submodel,
pathwise differentiability gives
\begin{align*}
\bbE\!\bigl[\phi_{\mathrm{eff}}^{\mathrm{CIM}}(O)s(O)\bigr]
&=\left.\frac{d}{dt}\psi(P_{O,t})\right|_{t=0}\\*
&=\bbE\!\bigl[\phi_{\mathrm{eff}}^{\mathrm{MAR}}(O)s(O)\bigr].
\end{align*}
The difference of the two canonical gradients is therefore orthogonal
to $\mathcal T$. Each canonical gradient belongs to $\mathcal T$, so
their difference also belongs to $\mathcal T$ and must be zero in
$L^2(P_{O,0})$. For a vector-valued target, apply this argument
componentwise. The efficient influence functions thus coincide almost
surely, and their common covariance is the common efficiency bound.
\end{proof}

\subsection{Proof of Theorem~\ref{thm:gaussian-tail-identification}}\label{app:gaussian-tail-proof}
\begin{proof}[Proof of Theorem~\ref{thm:gaussian-tail-identification}]
Suppose such a CIM law exists. Write its marginal as $N_p(\mu,\Sigma)$
and the true marginal as $N_p(\mu_0,\Sigma_0)$. Write
$\pi_{r,\theta}(u)=P'_{Y,R}(R=r\mid Y^{(r)}=u)$
for the candidate law's observed-pattern probabilities. Equality of the
observed complete-case subdistributions and the bound on a pattern
probability give
\[
f(y;\theta_0)\pi_{\mathbf1}(y)
=f(y;\theta)\pi_{\mathbf1,\theta}(y)
\leq f(y;\theta).
\]
By \textup{(\ref{eq:gaussian-tail-main})}, for every $\epsilon>0$,
\[
\log\frac{f(y;\theta_0)}{f(y;\theta)}
\leq \epsilon\|y\|^2-\log c_\epsilon
\quad\hbox{a.e. outside a bounded set}.
\]
The log ratio of the Gaussian densities has quadratic part
$-y^{\trans}(\Sigma_0^{-1}-\Sigma^{-1})y/2$. A negative eigenvalue of
$\Sigma_0^{-1}-\Sigma^{-1}$ would violate the displayed bound on a cone
of positive Lebesgue measure for a sufficiently small $\epsilon$.
Therefore $\Sigma_0^{-1}-\Sigma^{-1}\succeq0$, or equivalently
$D:=\Sigma-\Sigma_0\succeq0$.

Assume for contradiction that $(\mu,\Sigma)\ne(\mu_0,\Sigma_0)$.
We construct one direction $v$ along which every changed observed
Gaussian marginal has vanishing true-to-candidate density ratio, while
all unchanged observed subvectors remain fixed. Choose
\[
v=(\mu-\mu_0)+Dw.
\]
For any observed coordinate set $A$ with $D_{AA}\ne0$, the matrix
$\Sigma_{0,AA}^{-1}-\Sigma_{AA}^{-1}$ is nonzero and nonnegative definite,
and
\[
\operatorname{tr}\bigl\{
(\Sigma_{0,AA}^{-1}-\Sigma_{AA}^{-1})D_{AA}\bigr\}
=\|\Sigma_{AA}^{-1/2}D_{AA}\Sigma_{0,AA}^{-1/2}\|_F^2>0.
\]
Thus the affine set of $w$ for which
$v_A^{\trans}(\Sigma_{0,AA}^{-1}-\Sigma_{AA}^{-1})v_A=0$
is proper. There are finitely many patterns, so $w$ can be chosen
outside all these sets. For $D_{AA}=0$, positive semidefiniteness of
$D$ implies $D_{A,\{1,\ldots,p\}}=0$, and hence
$v_A=\mu_A-\mu_{0,A}$. It follows that, for every fixed $y$,
\[
\frac{f_r((y+tv)^{(r)};\theta_0)}
     {f_r((y+tv)^{(r)};\theta)}\longrightarrow0
\quad (t\longrightarrow\infty)
\]
whenever the two pattern marginals differ. For a covariance change the
negative quadratic term gives this limit. For a mean-only change the
linear coefficient is
$-(\mu_A-\mu_{0,A})^{\trans}\Sigma_{0,AA}^{-1}(\mu_A-\mu_{0,A})<0$.
If the two marginals agree, $v_A=0$ and their ratio is one.

Equality of the observed-data laws implies
\[
\pi_{r,\theta}(u)
=\pi_r(u)\frac{f_r(u;\theta_0)}{f_r(u;\theta)},
\qquad 0\leq\pi_r(u)\leq1.
\]
The CIM normalization of the candidate law holds Lebesgue-almost
everywhere. Outside a countable union of translated null sets, it holds
at $y+mv$ for every positive integer $m$. Let $m\to\infty$.
All terms with changed marginals vanish by the displayed tail limit; the
remaining terms do not vary along the line. Hence
\[
\sum_{r:\,P_{Y^{(r)},\theta}=P_{Y^{(r)},\theta_0}}
\pi_r(Y^{(r)})=1
\qquad P_{Y,\theta_0}\hbox{-a.s.}
\]
Taking expectation gives
$\sum_{r:\,P_{Y^{(r)},\theta}=P_{Y^{(r)},\theta_0}}\rho_r=1$.
The complete-case pattern is excluded from this sum when the two full
Gaussian marginals differ, yet \textup{(\ref{eq:gaussian-tail-main})}
implies $\rho_{\mathbf1}>0$. This is a contradiction.

Therefore $P'_Y=P_Y$. Equal observed and complete-data marginals imply
identical $\pi_r$, so the true law also satisfies CIM. \end{proof}

\subsection{Regularity conditions for the fitted test}
\begin{assumption}[Sufficient conditions for the fitted test]
\label{ass:testing}
The following conditions hold at the limiting parameter, denoted by
$\theta_0$ under the null and by $\theta^*$ under a fixed alternative.
\begin{enumerate}[(iii)]
\item[(i)] The observed units are independent and identically distributed.
The fixed kernel is measurable and bounded, its RKHS is separable, and all Hilbert-valued conditional means are strongly measurable. 
The model parameter has a fixed finite dimension and its limit is an interior point.
\item[(ii)] For a convex neighborhood of the limiting parameter, $\theta\mapsto\xi(O;\theta)$ is twice continuously Fr\'echet differentiable almost surely. 
The first and second derivatives have a square-integrable envelope in operator norm on that neighborhood.
\item[(iii)] The estimator has the expansion
\[
\sqrt n(\widehat\theta-\theta_0)
=\frac1{\sqrt n}\sum_{i=1}^n\operatorname{IF}_\theta(O_i)+o_p(1),
\quad
\bbE\operatorname{IF}_\theta=0,
\quad
\bbE\|\operatorname{IF}_\theta\|^2<\infty.
\]
Let $N_i^{(b)}$ be the number of times row $i$ is selected in bootstrap replicate $b$ of size $n$ drawn uniformly with replacement from the observed rows.
The refitted estimator obeys
\[
\sqrt n(\widehat\theta^{(b)}-\widehat\theta)
=\frac1{\sqrt n}\sum_{i=1}^n(N_i^{(b)}-1)
       \operatorname{IF}_\theta(O_i)+o_{P^*}(1)
\quad\text{in probability}.
\]
Here $P^*$ refers to bootstrap randomness conditional on the original sample. At a fixed alternative these same statements are required with $\theta_0$ replaced by the unique $\theta^*$.
\item[(iv)] Under $H_0$ in \eqref{eq:distributional-null}, the true complete-data parameter $\theta_0$ satisfies $g_{\cH}(\theta_0)=0$ and is consistently estimated as required by the preceding expansion.
For an asymptotically exact size assertion, the limiting squared Gaussian norm is nondegenerate and its $(1-\alpha)$ quantile is a uniquely determined continuity point.
\end{enumerate}
\end{assumption}

\label{app:testing}

Assumption~\ref{ass:testing} states the
conditions for the fitted test. Its expansion uses differentiable
empirical means and a conditional bootstrap central limit theorem
\citep{VanDerVaartWellner1996}. The following results give sufficient
conditions for the estimator and conditional-completion map.

\begin{proposition}[A sufficient M-estimation route]
\label{prop:direct-likelihood-if}
Suppose the selected original and bootstrap direct-likelihood stationary
points are consistent, the expected Hessian is nonsingular, the score is
square integrable, and uniform laws of large numbers hold for the Hessian
near the limiting parameter in both samples. Then the original and
bootstrap asymptotic linearity conditions in
Assumption~\ref{ass:testing}(iii) hold with
\begin{equation}
\operatorname{IF}_\theta(O)
=-\left[\bbE\left(
 \left.\frac{\partial^2\log f_R(Y^{(R)};\theta)}
      {\partial\theta\partial\theta^{\trans}}\right|_{\theta=\theta_0}\right)\right]^{-1}
 \left.\frac{\partial\log f_R(Y^{(R)};\theta)}{\partial\theta}\right|_{\theta=\theta_0}.
\label{eq:direct-if}
\end{equation}
The resulting variance is the observed-law sandwich variance.
\end{proposition}

\begin{proof}
Taylor expansion of the original and bootstrap score equations about the
limiting parameter, together with the stated consistency, Hessian, moment,
and uniform-convergence conditions, gives the two expansions in
Assumption~\ref{ass:testing}(iii) and the influence function in
\eqref{eq:direct-if}.
\end{proof}

Boundary fits, weak identification, and selection among full models
are not covered by these conditions.

\begin{corollary}[Gaussian full models]\label{cor:gaussian-regularity}
Suppose CIM holds and the complete-data marginal is a nondegenerate
Gaussian law. Parameterize the full model by its mean and covariance
on a compact set with covariance eigenvalues bounded above and away
from zero, and suppose the true parameter is interior. If
$\bbP(R=(1,\ldots,1)\mid Y)$ is bounded away from zero, then consistent
original and row-bootstrap maximizers of the direct likelihood satisfy
the asymptotic linearity in Assumption~\ref{ass:testing}. For a fixed
Gaussian kernel, the derivative-envelope conditions also hold.
Thus Theorem~\ref{prop:rkhs-limit} applies when its nondegenerate-quantile
condition is satisfied.
\end{corollary}

\begin{proof}
At the true law, the fixed probabilities $\pi_r$ satisfy the MAR
normalization for every Gaussian candidate on the common support.
The densities $\pi_r(u)f_r(u;\theta)$ therefore define an
observed-data parametric family. Its expected log likelihood differs
from the direct criterion by a term independent of $\theta$.
The positive complete-case probability identifies the Gaussian
parameter, and the observed score information is bounded below by a
positive multiple of the complete-data Gaussian information.
Gaussian scores and Hessians have polynomial envelopes on the compact
parameter set. These envelopes have finite moments under the Gaussian
marginal and yield the original and conditional-bootstrap uniform
laws. Proposition~\ref{prop:direct-likelihood-if} gives the linear
expansions. Differentiating Gaussian conditional densities twice
produces polynomials in observed and completed coordinates times those
densities. Boundedness of the kernel and uniformly bounded conditional
covariances give square-integrable polynomial envelopes for the
Hilbert-valued derivatives.
\end{proof}

\subsection{Remaining proofs}\label{app:theory}\label{app:theoretical-proofs}
\begin{proof}[Proof of Proposition~\ref{prop:three-pattern-counterexample}]
The assumption and iterated expectation give
$\bbE(H(Y)\mid Y_1)=0$. Hence the observed-pattern probabilities are
\[
\pi_{111}(Y)=\rho_{111},\qquad
\pi_{110}(Y_1,Y_2)=\rho_{110},\qquad
\pi_{100}(Y_1)=\rho_{100},
\]
and their sum is one. If $H$ depends on $Y_3$, the full-data
probabilities of the incomplete patterns depend on a variable missing
under those patterns, so MAR fails. For each pattern, integrating out the
missing coordinates removes the perturbation; the resulting observed-data
density is therefore $\rho_r f_r(y^{(r)};\theta_0)$, exactly as under the stated
MCAR mechanism.
\end{proof}

\begin{proof}[Proof of Lemma~\ref{lem:consistency}]
The expectation of the marginal likelihood ratio is
\begin{align*}
\bbE\!\left[\frac{f_R(Y^{(R)};\theta)}
                   {f_R(Y^{(R)};\theta_0)}\right]
&=\sum_{r\in\mathcal R}\int
 \pi_r(u)f_r(u;\theta)\,d\mu^{(r)}(u)\\
&=\int f(y;\theta)\sum_{r\in\mathcal R}
       \pi_r(y^{(r)})\,d\mu(y)=1.
\end{align*}
The last equality uses CIM and
$P_{Y,\theta}\ll P_{Y,\theta_0}$. Jensen's inequality yields
\[
\bar\ell(\theta)-\bar\ell(\theta_0)
=\bbE\log\frac{f_R(Y^{(R)};\theta)}
                 {f_R(Y^{(R)};\theta_0)}
\leq 0.
\]
For any fixed $\delta>0$, the separation assumption gives a positive gap
between $\bar\ell(\theta_0)$ and its supremum outside the $\delta$-ball. The
uniform law of large numbers makes the corresponding empirical gap
positive with probability tending to one. A maximizer cannot then lie
outside that ball. This proves consistency. The same argument permits
an approximate maximizer whose criterion error is $o_p(n)$.
\end{proof}

\begin{proof}[Proof of Theorem~\ref{thm:equivalence}]
Conditional on $R=r$, the observed subvector has density
$f_r(y^{(r)}\mid R=r)$ and the generated missing subvector has
conditional density $f(y^{(-r)}\mid y^{(r)};\theta_0)$.
Multiplication by $\rho_r$ gives the first expression in
\textup{(\ref{eq:pseudo-counterpart-law})}, and \textup{(\ref{eq:pi})} gives the second.
Comparison of \textup{(\ref{eq:pseudo-counterpart-law})} with
\textup{(\ref{eq:kano-counterpart-law})} proves the first equality in
\textup{(\ref{eq:pseudo-counterpart-identity})},
$\widetilde P_{Y,R,\theta_0}=(P_{Y,R})^{\mathrm{cp}}$; summing
\textup{(\ref{eq:pseudo-counterpart-law})} over the patterns and using
\textup{(\ref{eq:F})} proves the second,
$\widetilde f(y;\theta_0)=f(y;\theta_0)\{1+F(y)\}$.

Marginalizing \textup{(\ref{eq:pseudo-counterpart-identity})} over $R$ gives
\[
\widetilde P_{Y,\theta_0}=P_Y^{\mathrm{cp}}.
\]
Since $P_Y=P_{Y,\theta_0}$, the final equivalence of Theorem~\ref{thm:equivalence} follows immediately from
Theorem~\ref{thm:estimation-transfer}.
\end{proof}

\begin{proof}[Proof of Theorem~\ref{prop:rkhs-limit}]
\label{app:rkhs-sketch}
We first give the proof for fixed coordinates; the estimated-coordinate
case follows by the joint expansion in Section~\ref{supp:estimated-scales}.
Put $\dot\xi(O;\theta_0)=\partial\xi(O;\theta)/\partial\theta|_{\theta_0}$.
A Taylor expansion of the empirical kernel mean gives
\begin{align*}
\widehat g_{\cH}
={}&\frac1n\sum_{i=1}^n\xi(O_i;\theta_0)
 +\left(\frac1n\sum_{i=1}^n\dot\xi(O_i;\theta_0)\right)
                  (\widehat\theta-\theta_0)
 +O_p(\|\widehat\theta-\theta_0\|^2).
\end{align*}
The derivative envelope and finite parameter dimension yield a law of
large numbers in the corresponding finite product of Hilbert spaces.
The empirical derivative therefore converges in operator norm to
$\dot g_\theta=\bbE\dot\xi(O;\theta_0)$. Substitution of the estimator's
linear expansion proves \textup{(\ref{eq:nuisance-expansion})}. Under the null
$\zeta$ has mean zero and finite second moment, so the central limit
theorem in a separable Hilbert space gives the asserted Gaussian limit.
Its covariance is trace class since
$\operatorname{tr}\Var(\zeta)=\bbE\|\zeta\|_{\cH}^2<\infty$.
The spectral representation and the continuous mapping theorem give
\textup{(\ref{eq:weighted-chi-square})}.

Let $N_i^{(b)}$ count the selections of observed row $i$ in a bootstrap
sample. Taylor expansion of the original and bootstrap means about
$\theta_0$ and subtraction give
\[
\sqrt n(\widehat g_{\cH}^{(b)}-\widehat g_{\cH})
=\frac1{\sqrt n}\sum_{i=1}^n(N_i^{(b)}-1)\zeta(O_i)
 +o_{P^*}(1)
\quad\text{in probability}.
\]
The same derivative envelope controls the two Taylor remainders and the
bootstrap empirical derivative. Conditional bootstrap convergence follows from the empirical-process
bootstrap theorem of \citet{GineZinn1990}, applied to the unit-ball
linear functionals on the separable Hilbert space. For completeness,
it can also be seen by projecting on the first $J$
coordinates of an orthonormal basis. For fixed $J$, the usual
finite-dimensional bootstrap central limit theorem applies. For the
orthogonal tail the conditional second moment of the centred bootstrap
mean is bounded by its empirical squared tail norm. That norm converges
to its expectation, and the expectation tends to zero as $J\to\infty$
because $\bbE\|\zeta\|_{\cH}^2<\infty$. This proves conditional tightness
and the asserted Hilbert-space bootstrap limit. Squaring the norm and
using quantile continuity proves asymptotic level. An empirical
bootstrap quantile also converges when the number of independent
bootstrap resamples tends to infinity.

At a fixed alternative the same Taylor argument, centred by
$g_{\cH}(\theta^*)$, yields a finite-second-moment contribution
\[
\xi(O;\theta^*)-g_{\cH}(\theta^*)
 +\left.\frac{\partial g_{\cH}(\theta)}{\partial\theta}
  \right|_{\theta^*}\operatorname{IF}_{\theta^*}(O).
\]
It follows that $\widehat g_{\cH}\to g_{\cH}(\theta^*)$ and that the
centred refitted-bootstrap critical value is bounded in probability.
If $\Delta_{\cH}(\theta^*)>0$, the observed statistic grows linearly in
$n$, proving consistency. Under correct specification of the
full model and identification of CIM as in
Definition~\ref{def:cim-identification}, any non-CIM law in that
comparison class has positive
exact characteristic-kernel discrepancy at every candidate whose
observed-coordinate marginals dominate the corresponding observed-pattern
laws, provided that the CIM explanation constructed under zero discrepancy
belongs to the same comparison class, including any common design
restrictions. Otherwise the argument in Remark~\ref{rem:observational-equivalence-power} would provide a CIM
law in this class with the same observed distribution, contradicting
identification. In particular,
$\Delta_{\cH}(\theta^*)>0$ at a limiting candidate satisfying these
conditions. Without this separation, the power assertion still
applies to each fixed alternative with positive discrepancy; it need
not apply to the observationally equivalent alternatives in
Remark~\ref{rem:observational-equivalence-power}.

For a quadratic-mean differentiable observed-data path $P_{O,t}$ with score $s$,
$\bbE s=0$ and $\bbE s^2<\infty$, the local likelihood ratio at
$n^{-1/2}$ has the usual expansion
\[
\log\frac{dP_{O,n^{-1/2}}^{\otimes n}}{dP_{O,0}^{\otimes n}}
=\frac1{\sqrt n}\sum_{i=1}^n s(O_i)
 -\frac12\bbE s(O)^2+o_p(1).
\]
The joint central limit theorem for the Hilbert-valued sum of $\zeta$
and the scalar score sum, followed by Le Cam's third lemma, shifts the mean
of $\mathbb G_{\cH}$ by $\bbE(\zeta(O)s(O))$. Contiguity carries the
Taylor remainder and the bootstrap critical-value convergence to the
local sequence. This proves \textup{(\ref{eq:local-alternative})}; local power is
\[
\bbP\!\left(
\|\mathbb G_{\cH}+\bbE(\zeta(O)s(O))\|_{\cH}^2
>c_{1-\alpha}\right),
\]
at continuity points, where $c_{1-\alpha}$ is the null quantile. The
shift can vanish in directions indistinguishable after full-model
adaptation. 
\end{proof}

\begin{proof}[Proof of Theorem~\ref{thm:conditional-shape-identification}]
For any compatible CIM law, main-paper
Proposition~\ref{prop:exclusive-variable}
gives \textup{(\ref{eq:exclusive-block-main})}; the product-support
argument is in Section~\ref{supp:exclusive-variable-proof-examples}.
For the true law, the selected conditional density is instead
\[
f(v\mid x,R=r)=
\frac{\pi_r(x,v)}{\bbP(R=r\mid X=x)}f(v\mid x;\theta_0).
\]
A compatible CIM law must therefore represent the multiplier on the
right as a ratio of candidate to true conditional densities.

For Gaussian densities this ratio is $\exp(Av^2+Bv+C)$. Since the
selection probability is bounded, $A>0$ is impossible, and $A=0$
requires $B=0$. If $A<0$, the ratio tends to zero at both ends of
$\mathbb R$, which is incompatible with a positive nonconstant
monotone function. This proves the Gaussian case.

For any Student $t$ density, its ratio at $t$ and $-t$ tends to one
as $t\to\infty$, irrespective of its location, scale, and positive
finite degrees of freedom. A positive bounded nonconstant monotone
selection function has unequal two-sided limits, so the selected
density fails this tail balance. It cannot be a candidate Student $t$
density.

All contradictions occur on a set of covariates with positive
probability, proving separation of the observed laws.
\end{proof}

\begin{proof}[Proof of Proposition~\ref{prop:expfam-moment-matching}]
\label{app:moment-matching-proof}
For pattern $r$, integrating the complete-data density over its
missing coordinates and differentiating under the integral gives
\[
\nabla_\eta\log f_r(u;\eta)
=\frac{\int f_\eta(y)\{\tau(y)-\nabla\psi(\eta)\}\,dy^{(-r)}}
       {f_r(u;\eta)}
=\bbE_\eta\{\tau(Y)\mid Y^{(r)}=u\}-\nabla\psi(\eta).
\]
Here $y^{(r)}=u$ in the integral, and the conditional expectation
is under the full model, not the actual selection mechanism.
Regularity of the exponential family gives
$\nabla\psi(\eta)=\bbE_\eta\{\tau(Y)\}$.
Summing the displayed scores over the observed rows and using the
zero-score condition at $\widehat\eta$ proves
\textup{(\ref{eq:expfam-moment-matching})}.
For the Gaussian family, the sufficient-statistic vector consists of
$Y$ and the distinct entries of $YY^{\trans}$; their model means are
$\mu$ and $\Sigma+\mu\mu^{\trans}$. The argument is a finite-sample
identity of the fitting criterion, so it requires neither CIM nor
correct specification of the data-generating law. Penalized or
boundary fits need not satisfy it, and a numerical fit satisfies it
only to the accuracy of its score equations.
\end{proof}

\section{Calibration and computation of the fitted-kernel test}\label{supp:testing-details}\label{supp:other-topics}

\subsection{Multiplier calibration}\label{supp:multiplier}
For each original sample, estimate the first-order contribution from
Theorem~\ref{prop:rkhs-limit}{} of the main paper by
\begin{align*}
\widehat\zeta_i
&=\xi(O_i;\widehat\theta)
  +\widehat{\dot g}_\theta\widehat{\operatorname{IF}}_\theta(O_i),
\qquad
\widehat{\dot g}_\theta
 =\frac1n\sum_{i=1}^n
  \left.\frac{\partial\xi(O_i;\theta)}{\partial\theta}\right|_{\widehat\theta},
\\
\overline{\widehat\zeta}
&=\frac1n\sum_{i=1}^n\widehat\zeta_i.
\end{align*}
The $p$-value is computed from the $B$ multiplier draws exactly as in Algorithm~\ref{alg:cim-calibration} of the main paper, with the draws of the two multiplier types kept separate.
For estimated coordinate means and SDs, add the coordinate derivative
term in Section~\ref{supp:estimated-scales} before centring
$\widehat\zeta_i$. For fixed coordinates the displayed formula is
unchanged. The derivative holds the observed rows fixed. At an interior stationary
point of the direct likelihood, in the coordinates used for fitting, let
\[
\dot\ell_i(\theta)=\frac{\partial\log f_{R_i}(Y_i^{(R_i)};\theta)}{\partial\theta},
\qquad
\widehat A_\theta=\frac1n\sum_{i=1}^n
 \left.\frac{\partial\dot\ell_i(\theta)}{\partial\theta^{\trans}}\right|_{\widehat\theta}.
\]
When $\widehat A_\theta$ is nonsingular, use
\[
\widehat{\operatorname{IF}}_\theta(O_i)
=-\widehat A_\theta^{-1}\dot\ell_i(\widehat\theta).
\]
The minus sign corresponds to defining $\widehat A_\theta$ as the
Hessian of the log likelihood. A linear solve evaluates this expression.
Both the score and Hessian differentiate the observed-data likelihood.
For independent weights $e_i$ with mean zero and variance one, define
\begin{equation}
T_{n,\cH}^{\mathrm{mult}}
 =\left\|\frac1{\sqrt n}\sum_{i=1}^n
 e_i(\widehat\zeta_i-\overline{\widehat\zeta})\right\|_{\cH}^2.
\label{eq:multiplier-statistic}
\end{equation}
Standard Gaussian weights and Rademacher weights, taking $-1$ and $1$
with equal probability, define MULT\_G and MULT\_R, respectively.
Both use the corrected, centred contributions in
\eqref{eq:multiplier-statistic} and calibrate the same original fitted
statistic $T_{n,\cH}$ with separate multiplier calibration draws.
This is a fitted-parameter version of multiplier calibration for
empirical processes and goodness-of-fit tests
\citep{VanDerVaartWellner1996,KojadinovicYan2012}.

\begin{theorem}[Multiplier calibration for the fitted statistic]
\label{thm:multiplier}
Suppose the null expansion in Theorem~\ref{prop:rkhs-limit}{} of the main paper holds with
$\bbE\|\zeta(O)\|_{\cH}^2<\infty$, and suppose
\begin{equation}
\frac1n\sum_{i=1}^n
\|\widehat\zeta_i-\zeta(O_i)\|_{\cH}^2=o_p(1).
\label{eq:multiplier-l2}
\end{equation}
For Gaussian or Rademacher weights, with draws independent conditional
on the observations,
the conditional law of the vector inside the norm in
\eqref{eq:multiplier-statistic} converges in probability to the law of
$\mathbb G_{\cH}$ from that theorem. With the same nondegenerate,
continuous-quantile condition, multiplier calibration is asymptotically
exact at level $\alpha$. At a fixed alternative with positive fitted
discrepancy, the resulting test is consistent if the estimated centred
contributions have bounded empirical second moment in probability.
\end{theorem}

\begin{proof}
First replace the estimated contributions by
$\zeta(O_i)-n^{-1}\sum_j\zeta(O_j)$. Finite-dimensional projections
satisfy the conditional multiplier central limit theorem. The
conditional mean squared norm of the omitted Hilbert-space tail is
bounded by the empirical squared tail norm. Its limit decreases to
zero with the projection dimension because $\zeta$ has a finite second
moment. This establishes the Hilbert-space result, equivalently the
multiplier theorem applied to the unit-ball linear functionals
\citep{VanDerVaartWellner1996}.
Conditional on the data, the mean squared norm of the difference made
by using $\widehat\zeta_i$ is at most
$n^{-1}\sum_i\|\widehat\zeta_i-\zeta(O_i)\|_{\cH}^2$.
Equation~\eqref{eq:multiplier-l2} therefore transfers the limit.
Continuity of the squared norm and of the limiting quantile proves
level control. At a fixed alternative, the observed statistic grows
linearly in $n$, while the conditional expectation of
$T_{n,\cH}^{\mathrm{mult}}$ is the empirical centred second moment.
Its boundedness makes the critical value bounded in probability,
proving consistency.
\end{proof}

Condition~\eqref{eq:multiplier-l2} includes estimation of the derivative
and influence function as well as kernel approximation. For example,
it follows from empirical mean-square consistency of the row terms and
influence functions, operator-norm consistency of the derivative, and
bounded empirical second moments of the influence functions. The
method removes repeated nonlinear optimization from calibration.
The selected calibration is stated alongside each $p$-value.
For Monte Carlo calibration, the asymptotic level assertion uses
$B\to\infty$.

\subsection{Estimated coordinate locations and scales}
\label{supp:estimated-scales}
Let $Y$ denote the measurements after any fixed scientific transformations.
For $\eta=(a,s)$ with $s_j>0$, write $L_\eta(y)=D_s^{-1}(y-a)$,
$D_s=\operatorname{diag}(s)$. Hold the base kernel $k_0$ and its RKHS
$\cH_0$ fixed and put $\phi_\eta(y)=k_0(L_\eta(y),\cdot)$. Parameterize
the full model on the original $Y$ coordinates, before centring or
scaling. For an observed row with $R=r$, define
\[
 \xi(O;\theta,\eta)=\bbE_\theta\{\phi_\eta(Y)\mid Y^{(r)}\}
                -\bbE_\theta\{\phi_\eta(Y)\},\qquad
 g(\theta,\eta)=\bbE\{\xi(O;\theta,\eta)\},
\]
where the outer expectation holds the observed law fixed. If $P_n$ and
$P_n^{(b)}$ are the empirical and resampled row laws, set
$\widehat g=P_n\xi(\cdot;\widehat\theta,\widehat\eta)$ and
$\widehat g^{(b)}=P_n^{(b)}\xi(\cdot;\widehat\theta^{(b)},\widehat\eta^{(b)})$.
The statistic and its bootstrap draw are $n\|\widehat g\|^2_{\cH_0}$
and $n\|\widehat g^{(b)}-\widehat g\|^2_{\cH_0}$. A Gaussian
completion with mean $m$ and covariance $V$ has transformed parameters
$L_\eta(m)$ and $D_s^{-1}VD_s^{-1}$. Use the respective fitted
transformations on the two sides of every inner product in
Section~\ref{app:kernel-calculations}, including the resample--original
cross term.

For each variable $j$, declare a subset where it is observed, with row
indicator $I_j$ and probability $\rho_j>0$. Set
$a_j=\bbE(Y_j\mid I_j=1)$ and $s_j^2=\Var(Y_j\mid I_j=1)>0$. If
the subset distribution has finite fourth moments, the mean and sample
SD have influence functions
\[
 \operatorname{IF}_{a_j}=\frac{I_j}{\rho_j}(Y_j-a_j),\qquad
 \operatorname{IF}_{s_j}=\frac{I_j}{2\rho_js_j}\{(Y_j-a_j)^2-s_j^2\},
\]
both zero off the subset. In the bivariate analysis $I_j=I\{R=11\}$.
With $m_j=\sum_i I_{ij}$ and $\widehat\rho_j=m_j/n$, the plug-in
mean influence uses $I_{ij}(Y_{ij}-\widehat a_j)/\widehat\rho_j$.
For the SD with denominator $m_j-1$ we use the empirically centred
version
\[
 \widehat{\operatorname{IF}}_{s_j}(O_i)=
 \frac{I_{ij}\{(Y_{ij}-\widehat a_j)^2-\widehat s_j^2\}
              +\widehat s_j^2/n}
      {2\widehat s_j(\widehat\rho_j-1/n)}.
\]
These summarize an observed subset, not necessarily the full-data population.

Let $(\theta_*,\eta_*)$ be the limits of the fit and coordinate rule at
a regular observed law, $g_*=g(\theta_*,\eta_*)$, and $G_\theta,G_\eta$
the partial derivatives of $g$ there. Extend Assumption~\ref{ass:testing}
to the joint parameter: the two estimators have joint asymptotically
linear original and row-bootstrap expansions; the population map is
differentiable and the centred empirical and bootstrap processes are
locally stochastically equicontinuous in $(\theta,\eta)$. For example,
the smooth finite-dimensional case with the derivative envelopes of
that assumption and scales bounded away from zero in a neighbourhood
of $\eta_*$ suffices. Adding and subtracting
$\bbE\xi(O;\widehat\theta,\widehat\eta)$, equicontinuity handles the
empirical-process remainder and differentiation of the population term gives
\[
 \sqrt n(\widehat g-g_*)=n^{-1/2}\sum_i\zeta_*(O_i)+o_p(1),\qquad
 \zeta_*=\xi(\cdot;\theta_*,\eta_*)-g_*+
              G_\theta\operatorname{IF}_\theta+G_\eta\operatorname{IF}_\eta .
\]
The same expansion conditional on the original rows, with bootstrap
counts minus one as weights, gives the centred refitted-bootstrap limit.

Under the joint null, completion preserves the law of $Y$, hence
$g(\theta_0,\eta)=0$ for \emph{every} $\eta$. Thus $G_\eta=0$ at fixed
original-scale $\theta_0$. This also holds for the conditional discrepancy
of Section~\ref{supp:conditional-reference-centering} when the completed
and model laws agree given the always-observed covariates. The null
limit and the local shift in Theorem~\ref{prop:rkhs-limit} therefore
use $\zeta_0=\xi(\cdot;\theta_0,\eta_0)+G_\theta\operatorname{IF}_\theta$;
the local shift can be zero. At a regular fixed alternative,
$T_{n,\cH}/n\to_p\|g_*\|^2_{\cH_0}$ and the bootstrap critical value is
bounded in probability. A characteristic base kernel and invertible
$L_{\eta_*}$ preserve the zero set of the discrepancy, so the test is
consistent when the completed and model laws differ at the limiting fit.

For multiplier calibration we retain the empirical terms
$\widehat G_\theta\widehat{\operatorname{IF}}_\theta+
\widehat G_\eta\widehat{\operatorname{IF}}_\eta$, where
$\widehat G_u=P_n\partial_u\xi(\cdot;\widehat\theta,\widehat\eta)$, in
the centred contributions of Section~\ref{supp:multiplier}. The empirical
coordinate derivative need not vanish, and its population counterpart may
be nonzero under alternatives. The qualification about fixed original-scale
parameters matters: if $\beta=B(\theta,\eta)$ denotes the standardized
model parameter and $\theta=H(\beta,\eta)$ its inverse in $\theta$, then
at fixed $\beta$ the coordinate derivative is
$G_\eta+G_\theta\partial_\eta H$. Since
$\operatorname{IF}_\beta=(\partial_\theta B)\operatorname{IF}_\theta+
(\partial_\eta B)\operatorname{IF}_\eta$, the chain rule recovers the
same total contribution in either parameterization. Dropping the
coordinate term at fixed standardized parameters is not generally valid.

For the Gaussian-kernel mixture and an equivariant Gaussian fit, the
finite-sample rule also respects changes of measurement origin and units.
After $Y_j'=c_j+d_jY_j$ ($d_j\ne0$), the standardized values in both
the original sample and every resample change only by the same coordinate
signs, leaving all Gaussian-kernel inner products unchanged. This requires
the same eligible rows and concerns changes made after the declared
scientific transformations. With scaling alone, a change of origin
adds different offsets to embeddings if the original and resampled SDs
differ; their cross inner product can then change. By contrast, within
a single sample the common translation preserves the statistic itself.
Nonpositive scales, data-driven selection among coordinate rules and
nonregular fits are not covered.

\subsection{Common-covariate centring}\label{supp:centering}
\label{app:centering-proof}\label{supp:centering-computation}

We first study signal preservation and oracle variance.
Let $C=Y_{\cC}$ contain the coordinates observed under every pattern.
For a fixed square-integrable feature vector
$\Phi_K=(\phi_1,\ldots,\phi_K)^{\trans}$, write
$m_r(Y^{(r)};\theta)=\bbE_\theta[\Phi_K(Y)\mid Y^{(r)}]$,
$\mu(\theta)=\bbE_\theta[\Phi_K(Y)]$, and
$g(C;\theta)=\bbE_\theta[\Phi_K(Y)\mid C]$. Define
\begin{align*}
\xi^G(\theta)&=m_{R}(Y^{(R)};\theta)-\mu(\theta),
\\
\xi^{CC}(\theta)&=m_{R}(Y^{(R)};\theta)-g(C;\theta).
\end{align*}
For a correct complete-data marginal, irrespective of CIM,
\begin{equation*}
\bbE[F(Y)\mid C]=0,\qquad
\bbE_{\theta_0}[(\phi(Y)-\bbE_{\theta_0}[\phi(Y)\mid C])F(Y)]
=\bbE_{\theta_0}[\phi(Y)F(Y)].
\end{equation*}
Thus
\begin{equation*}
\bbE[\xi^G(\theta_0)]=\bbE[\xi^{CC}(\theta_0)]=:a_K.
\end{equation*}
\begin{proposition}[Signal-preserving variance reduction]\label{prop:variance-reduction}
Under CIM and a correct full-model marginal, with $\theta_0$ fixed,
\begin{align*}
\Omega_G&=\Omega_{CC}+\Var[g(C;\theta_0)],\qquad
\Omega_q=\Var[\xi^q(\theta_0)],\quad q\in\{G,CC\},
\\
\Omega_{CC}&\preceq\Omega_G.
\end{align*}
For every square-integrable $C$-measurable vector $a(C)$,
\begin{equation*}
\Var[m_{R}(Y^{(R)};\theta_0)-a(C)]
=\Omega_{CC}+\Var[g(C;\theta_0)-a(C)]\succeq\Omega_{CC}.
\end{equation*}
\end{proposition}
\begin{proof}
Under CIM, completion preserves $C$ and has marginal $P_{Y,\theta_0}$,
so $\bbE[m_{R}(Y^{(R)};\theta_0)\mid C]=g(C;\theta_0)$.
Conditional-variance decomposition gives the variance identities.
Independently of CIM, iterated conditioning gives
$\bbE[F(Y)\mid C]=\sum_{r\in\mathcal R}\bbP(R=r\mid C)-1=0$,
which proves the signal identity.
\end{proof}
For $q\in\{G,CC\}$, let
\begin{equation*}
\widehat a^q=\frac1n\sum_i\xi_i^q(\theta_0),\qquad
T_{n,K}^q=n(\widehat a^q)^{\trans}(\widehat\Omega_q)^+\widehat a^q.
\end{equation*}
For fixed $K$, positive-definite $\Omega_q$, and consistent covariance
estimates, a local sequence retaining the correct complete-data marginal
with $\sqrt n\,a_K\to\delta$ and a central limit theorem yields
\begin{equation*}
T_{n,K}^q\overset{d}{\longrightarrow}\chi_K^2(\lambda_q),\qquad
\lambda_q=\delta^{\trans}\Omega_q^{-1}\delta,\qquad
\lambda_{CC}\geq\lambda_G.
\end{equation*}
This ordering concerns known parameters and fixed features in a Wald
test; fitted parameters and the full RKHS norm have different
first-order contributions.

For the fitted statistic, let $g_k(C;\theta)=\bbE_\theta[k(Y,\cdot)\mid C]$ and define
\begin{align*}
\xi^{CC}_{\cH}(O;\theta)&=m_{R,k}(Y^{(R)};\theta)-g_k(C;\theta),\\
\widehat g^{CC}_{\cH}&=\frac1n\sum_i\xi^{CC}_{\cH}(O_i;\widehat\theta),
\qquad T^{CC}_{n,\cH}=n\|\widehat g^{CC}_{\cH}\|_{\cH}^2.
\end{align*}
The fitted expansion and calibration in
Theorem~\ref{prop:rkhs-limit} of the main paper apply under its
conditions for this contribution, with
\begin{equation*}
\zeta^{CC}(O)=\xi^{CC}_{\cH}(O;\theta_0)
+\left.\bbE\!\left[\frac{\partial\xi^{CC}_{\cH}(O;\theta)}{\partial\theta}\right]
\right|_{\theta=\theta_0}\operatorname{IF}_\theta(O).
\end{equation*}
Each bootstrap refit recomputes both conditional means and uses its own
centred statistic
$n\|\widehat g^{CC,(b)}_{\cH}-\widehat g^{CC}_{\cH}\|_{\cH}^2$ and critical
value. Rows observing only $C$ contribute zero; normalization uses the
original sample size $n$.

For $g_k$, conditional integration holds $C$ fixed and generates every
other coordinate, including those observed in the original row. The
integration formulas and approximation requirements are given in
Section~\ref{app:gaussian-kernel-calculation} \citep{Kim2011,SangKimLee2022}. For the Gaussian kernel,
\[
[g_{k_\sigma}(c;\theta)](x)=\kappa_\sigma(a_{\cC}(c;\theta),V_{\cC}(\theta);x,0).
\]
Writing $a_i=a_{R_i}(Y_i^{(R_i)};\theta)$,
$V_i=V_{R_i}(\theta)$, $b_i=a_{\cC}(C_i;\theta)$, and
$W=V_{\cC}(\theta)$ gives
\begin{align*}
\langle\xi^{CC}_{\cH}(O_i;\theta),\xi^{CC}_{\cH}(O_j;\theta)\rangle_{\cH}
={}&\kappa_\sigma(a_i,V_i;a_j,V_j)-\kappa_\sigma(a_i,V_i;b_j,W)\\
&-\kappa_\sigma(b_i,W;a_j,V_j)+\kappa_\sigma(b_i,W;b_j,W).
\end{align*}
The formula also gives cross inner products at different fitted
parameters and feature means for $\phi_j(Y)=k_\sigma(Y,v_j)$.
For the multiscale kernel, average the inner products at
bandwidths $\sigma\in\{1,2,4\}$. Under model misspecification, conditional
centring can remove discrepancies in the marginal distribution of $C$;
signal preservation applies to a correct full-model marginal.

\subsection{Conditional models with an unspecified covariate marginal}
\label{supp:conditional-reference-centering}
When only $X\mid C$ is modelled, write the candidate complete-data law as
$P_C(dc)f_\theta(x\mid c)\,dx$, with $P_C$ unspecified and $C$ observed
in every row. For the product kernel, let
\[
m_{r,k}(c,u;\theta)
=\bbE_\theta\{k((c,X),\cdot)\mid X^{(r)}=u,C=c\},
\qquad
g_k(c;\theta)=\bbE_\theta\{k((c,X),\cdot)\mid C=c\};
\]
the row contribution is the conditional residual
$m_{r,k}(C,X^{(r)};\theta)-g_k(C;\theta)$, which coincides with
\eqref{eq:support2-conditional-contribution}. Plugging the empirical
covariate distribution into the global reference embedding yields
exactly the same sample average as conditional centring,
$n^{-1}\sum_i m_{R_i,k}(C_i,X_i^{(R_i)};\theta)-n^{-1}\sum_i g_k(C_i;\theta)
=n^{-1}\sum_i\xi_C(O_i;\theta)$, so the conditional analysis involves no
separate G/CC comparison. Under the joint
conditional-model/CIM null, completion preserves $C$ and the conditional
complete-data law, so the residual has conditional mean zero given $C$.
The fitted-parameter expansion and row-bootstrap calibration of
Theorem~\ref{prop:rkhs-limit} therefore apply with the influence
function of the conditional-likelihood fit, under the same
asymptotic-linearity, differentiability and bootstrap-regularity
conditions, with $\theta$ comprising every regression coefficient and
every free common-covariance parameter. The marginal law $P_C$ is
unrestricted.

\subsection{Completed sufficient-statistic matching}
We compare completed and model distributions through expectations of
their features. Throughout this subsection, expectations without a
parameter subscript use the observed law $P_O$ and the indicated
conditional completion, whereas $\bbE_\theta$ denotes expectation under
$P_{Y,\theta}$. The density identity \eqref{eq:pseudo-counterpart-identity} gives,
for any $P_{Y,\theta_0}$-integrable $\phi$,
\begin{equation*}
\bbE(\phi(\widetilde Y_{\theta_0}))
-\bbE_{\theta_0}(\phi(Y))
=\bbE_{\theta_0}(\phi(Y)F(Y)).
\end{equation*}
The right-hand side shows which CIM departures a particular feature can detect. 
The following proposition is not new: it is the Sundberg formula for the
observed-data score of an exponential family
\citep{Sundberg1974,DempsterLairdRubin1977}, the self-consistency
property underlying the EM algorithm, restated in the present notation
because it determines which discrepancies the fitted test can detect.

\begin{proposition}[Completed sufficient-statistic matching]
\label{prop:expfam-moment-matching}
Let $f_\eta(y)=h(y)\exp\{\eta^{\trans}\tau(y)-\psi(\eta)\}$ be a full
regular exponential family with common support. If differentiation
under marginalization is valid and $\widehat\eta$ is an interior
stationary point of the unpenalized direct log likelihood, then,
using the observed patterns $R_i$,
\begin{equation}
\frac1n\sum_{i=1}^n
 \bbE_{\widehat\eta}\{\tau(Y)\mid Y^{(R_i)}=Y_i^{(R_i)}\}
 =\bbE_{\widehat\eta}\{\tau(Y)\}.
\label{eq:expfam-moment-matching}
\end{equation}
This identity does not require CIM or correct specification of the
full-model family.
\end{proposition}
For completeness, the short derivation is given in Section~\ref{app:theoretical-proofs}.

\subsection{Exact Gaussian kernel calculations}\label{app:kernel-calculations}
\label{app:gaussian-kernel-calculation}

For $Y^{(r)}=u$, fractional imputation draws missing components
$V_1,\ldots,V_M$ from a proposal density $q$ covering the model
conditional support \citep{Kim2011,SangKimLee2022}. Let $\widetilde y_b$ retain
$u$ and complete the remaining coordinates with $V_b$. Then
\begin{align*}
w_b(\theta)
&=\frac{f(V_b\mid u;\theta)/q(V_b)}
{\sum_{a=1}^M f(V_a\mid u;\theta)/q(V_a)},
\\
m_{r,k}(u;\theta)
&\approx\sum_{b=1}^M w_b(\theta)k(\widetilde y_b,\cdot).
\end{align*}
Direct conditional draws give $w_b=1/M$.
Let $e_n=\widehat g_{\cH}^{\mathrm{app}}-\widehat g_{\cH}$ and
$e_n^{(b)}=\widehat g_{\cH}^{\mathrm{app},(b)}-\widehat g_{\cH}^{(b)}$
denote the numerical errors in the full fitted embeddings, including
their reference terms. The limits in
Theorem~\ref{prop:rkhs-limit} are preserved if
\[
\sqrt n\,\|e_n\|_{\cH}=o_p(1),
\qquad
\sqrt n\,\|e_n^{(b)}-e_n\|_{\cH}=o_{P^*}(1)
\quad\text{in probability}.
\]
Here $P^*$ includes bootstrap and numerical randomness, conditional on
the original data and numerical calculation. Multiplier calibration
additionally requires the approximated contributions to satisfy
\eqref{eq:multiplier-l2}.

For Gaussian models, these integrations have closed forms. Let
$Y\sim N_p(\mu,\Sigma)$ with $\Sigma$ positive definite, and let
$k_\sigma(x,y)=\exp(-\|x-y\|^2/(2\sigma^2))$ with $\sigma>0$. For a coordinate set $A$, write
$a_A(u;\theta)$ and $V_A(\theta)$ for the full $p$-dimensional mean and
covariance of $Y\mid Y_A=u$. The observed entries have zero conditional
variance, and
\begin{align*}
(a_A)_A&=u,
& (a_A)_{A^c}
&=\mu_{A^c}+\Sigma_{A^c A}\Sigma_{AA}^{-1}(u-\mu_A),
\\
(V_A)_{A^c A^c}
&=\Sigma_{A^c A^c}
-\Sigma_{A^c A}\Sigma_{AA}^{-1}\Sigma_{A A^c},
& (V_A)_{A,\{1,\ldots,p\}}&=0.
\end{align*}
The corresponding columns are zero. For $A=\varnothing$, take
$(a_A,V_A)=(\mu,\Sigma)$; for all coordinates observed, take $(u,0)$.
For pattern $r$, write $a_r(u;\theta)$ and $V_r(\theta)$
for these conditional parameters given $Y^{(r)}=u$.

For independent $Y\sim N_p(a,V)$ and $Y'\sim N_p(b,W)$, define
$\kappa_\sigma(a,V;b,W)=\bbE(k_\sigma(Y,Y'))$. Completing the square in
$Y-Y'\sim N_p(a-b,V+W)$ gives
\begin{equation}
\kappa_\sigma(a,V;b,W)
=\left|I_p+\frac{V+W}{\sigma^2}\right|^{-1/2}
\exp\!\left(-\frac12(a-b)^{\trans}
(\sigma^2I_p+V+W)^{-1}(a-b)\right).
\label{eq:gaussian-kernel-expectation}
\end{equation}
Here $I_p$ is the $p\times p$ identity matrix. The formula remains
valid for singular nonnegative-definite $V,W$, since
$\sigma^2I_p+V+W$ is positive definite. For $x\in\mathbb R^p$,
\begin{align*}
[m_{r,k_\sigma}(u;\theta)](x)
&=\kappa_\sigma(a_r(u;\theta),V_r(\theta);x,0),
\\
[\mu_{k_\sigma}(\theta)](x)
&=\kappa_\sigma(\mu,\Sigma;x,0).
\end{align*}

For a fixed fitted parameter, abbreviate
$a_i=a_{R_i}(Y_i^{(R_i)};\theta)$ and
$V_i=V_{R_i}(\theta)$. The row inner product in
\textup{(\ref{eq:rkhs-statistic})} is
\begin{align*}
&\langle\xi(O_i;\theta),\xi(O_j;\theta)\rangle_{\cH}
\\
&\quad=\kappa_\sigma(a_i,V_i;a_j,V_j)
-\kappa_\sigma(a_i,V_i;\mu,\Sigma)
-\kappa_\sigma(\mu,\Sigma;a_j,V_j)
+\kappa_\sigma(\mu,\Sigma;\mu,\Sigma).
\end{align*}
Summing and dividing by $n$ gives the statistic. The same formula applies
to cross inner products at different fitted parameters in the bootstrap.
Even for $i=j$, the two completed vectors are conditionally independent;
the diagonal is not generally $k_\sigma(y,y)=1$.

\section{Further identification results}\label{supp:identification-separation}\label{supp:additional-identification}\label{supp:continuous-identification}

A sufficient condition for the complete-case tail bound
\eqref{eq:gaussian-tail-main} of the main paper is uniform
complete-case positivity, $\bbP(R=\mathbf1\mid Y=y)\geq\epsilon_0>0$.
More generally, for a multinomial-logistic mechanism
$\bbP(R=r\mid Y=y)=\exp(a_r+b_r^{\trans}y)/\sum_{s\in\mathcal R}\exp(a_s+b_s^{\trans}y)$,
one has $-\log\bbP(R=\mathbf1\mid Y=y)\leq K(1+\|y\|)$ for a finite $K$,
which implies \eqref{eq:gaussian-tail-main}; the same argument
covers complete-case probabilities with polynomial tail decay.

\subsection{Conditional restrictions and non-Gaussian families}
Theorem~\ref{thm:conditional-shape-identification}{} of the main paper applies without complete cases. Let $Y=(Y_1,Y_2,Y_3)$ be Gaussian or Student $t$, retain
only patterns $110$ and $011$, and assign pattern $110$ with probability
$\expit(\alpha+\beta Y_1)$. For $\beta\ne0$, take $V=Y_1$ and $X=Y_2$
in Theorem~\ref{thm:conditional-shape-identification}; for $\beta=0$
the mechanism is MCAR. Thus CIM is identified in this class of missing-data mechanisms without a complete-case pattern. The model restriction is on the
conditional family of $V$ given $X$; the marginal law of $X$ need not
be specified.

\label{supp:conditional-identification}

\begin{proposition}[Identification by a positive tail limit]
\label{prop:positive-tail-limit}
Under the support and exclusive-coordinate conditions of Theorem~\ref{thm:conditional-shape-identification}{}
of the main paper, suppose the true and candidate conditional laws
$V\mid X=x$ are all nondegenerate Gaussian, or all Student $t$ with
arbitrary positive finite degrees of freedom. On a positive-probability
set of $x$, assume $\pi_r(x,\cdot)$ is positive and not almost
everywhere constant. The identification conclusion holds if, in at least one tail,
\begin{equation}
\pi_r(x,v)=c(x)+o(|v|^{-2}),\qquad c(x)>0.
\label{eq:shape-plateau-main}
\end{equation}
For Gaussian conditionals, convergence to a positive limit alone
suffices. No CIM law in the stated family then has the true
observed-data distribution.
\end{proposition}
\begin{proof}
Proposition~\ref{prop:exclusive-variable} of the main paper and
Bayes' formula require selection to be proportional to a
candidate-to-true conditional density ratio. For Gaussian densities
this ratio is $\exp(Av^2+Bv+C)$; a finite positive limit in one tail
forces $A=B=0$. Normalization then forces equal conditional densities
and constant selection, a contradiction.

For \eqref{eq:shape-plateau-main}, a positive finite limit of a Student
$t$ density ratio first forces equal conditional degrees of freedom,
denoted by $\lambda$. With true location and scale $m_0,s_0$ and
candidate values $m,s$, expansion of the log ratio in the specified
tail gives
\begin{align*}
\log\frac{f(v;m,s,\lambda)}{f(v;m_0,s_0,\lambda)}
={}&C+\frac{(\lambda+1)(m-m_0)}{v}\\
&-\frac{\lambda+1}{2v^2}
\{\lambda(s^2-s_0^2)-(m^2-m_0^2)\}+O(|v|^{-3}).
\end{align*}
The little-$o$ condition forces $m=m_0$ and then $s=s_0$. Selection
would be constant. All contradictions occur on a set of covariates
with positive probability, proving separation of the observed laws.
\end{proof}
Unlike Theorem~\ref{thm:conditional-shape-identification}, the
proposition does not require $\pi_r(x,\cdot)$ to be monotone; for
example, $\pi_r(x,v)=1/2+\exp(-v^2)/4$ is nonmonotone but satisfies
\eqref{eq:shape-plateau-main} in both tails.

\begin{proposition}[Identification by bounded conditional density ratios]
\label{prop:conditional-ratio-separation}
Under the pattern and support conditions of
Proposition~\ref{prop:exclusive-variable} of the main paper, suppose the true
complete-data marginal belongs to $\mathcal M$. Assume that, for almost
every $x$, any two conditional densities $f(v\mid x;\theta)$ and
$f(v\mid x;\theta')$ in the induced family satisfy
\begin{equation}
0<c\leq\frac{f(v\mid x;\theta')}{f(v\mid x;\theta)}\leq C<\infty
\quad\hbox{a.e. in }v
\quad\Longrightarrow\quad
f(\cdot\mid x;\theta')=f(\cdot\mid x;\theta).
\label{eq:conditional-density-rigidity}
\end{equation}
The constants can depend on $x,\theta,\theta'$. Suppose the true
$\pi_r(x,v)$ is bounded away from zero in $v$ for almost every $x$,
and is nonconstant in $v$ on a set of $x$ values of positive probability.
Then no CIM law with complete-data marginal in $\mathcal M$ has the
same observed-data distribution as the true law. For a characteristic
kernel, $\Delta_{\cH}(\theta)>0$ at every candidate whose observed-coordinate
marginals dominate the corresponding observed-pattern laws.
\end{proposition}

\begin{proof}
For the true law, Bayes' formula yields
\[
\frac{f(v\mid x,R=r)}{f(v\mid x;\theta_0)}
=\frac{\pi_r(x,v)}{\bbP(R=r\mid X=x)}.
\]
The ratio is bounded above and away from zero as a function of $v$.
An observationally equivalent CIM law would, by
Proposition~\ref{prop:exclusive-variable}, represent the
numerator as $f(v\mid x;\theta)$ for some candidate in $\mathcal M$.
Condition~\eqref{eq:conditional-density-rigidity} would force that
conditional density to equal the true one and hence force
$\pi_r(x,v)$ to be constant in $v$. This contradicts the stated
nonconstancy on a positive-probability set. The argument in Remark~\ref{rem:observational-equivalence-power} and the characteristic
property then give the discrepancy conclusion.
\end{proof}

Nondegenerate multivariate Gaussian conditional families satisfy
\eqref{eq:conditional-density-rigidity}: their log density ratio is a
quadratic polynomial, bounded in both directions only when constant.
This allows arbitrary conditional means and positive-definite
conditional covariances as functions of $X$; the marginal distribution
of $X$ need not be Gaussian. For scalar $V$, Poisson families with
positive means and gamma families with positive shape and rate also
satisfy the condition. Their log ratios are, respectively, affine in
$v$ and of the form $a\log v+bv+c$, and boundedness in both directions
on their full supports forces equality of the distributions. More
generally, a conditional exponential family has this property when
no nonconstant linear combination of its sufficient statistics is
bounded above and below on the conditional support.

In the ordinary response setting, where $X$ is always observed and
$V$ is either fully observed or entirely missing, CIM is equivalent to
$\pi_{\mathbf1}(X,V)=\bbP(R=\mathbf1\mid X)$. The preceding proposition therefore
identifies CIM throughout the model class with conditional family
\eqref{eq:conditional-density-rigidity} and
$\pi_{\mathbf1}(x,v)\geq c(x)>0$. For more general patterns the same argument
separates every alternative satisfying its nonconstancy condition.

\subsection{Gaussian alternatives with an exclusive variable}
\label{supp:original-gaussian-separation}

We construct a distributional departure that leaves the Gaussian fit unchanged.
For the four-variable Gaussian law of Experiment~A, take
$C=Y_1$ and $U=(Y_4-0.5Y_3)/\sqrt{0.75}$, so $U$ is standard normal
and independent of $(C,Y_2,Y_3)$. Starting with the four equiprobable
observation patterns of A0, transfer $\delta H(U)$ from $1000$ to $1111$,
leaving the other probabilities unchanged, where $0<\delta\leq1/4$ and
\[
 H(u)=\frac{\sin u-\tfrac12 e^{3/2}\sin(2u)}
              {1+\tfrac12 e^{3/2}} .
\]
The continuous periodic function $H$ has $\sup_u|H(u)|<1$: equality in
the triangle bound would require $|\sin u|=|\sin(2u)|=1$. Thus the
complete-case probability is bounded away from zero. Oddness and the
normal identity $\bbE\{U\sin(tU)\}=t\exp(-t^2/2)$ give
$\bbE\{H(U)\}=\bbE\{UH(U)\}=\bbE\{U^2H(U)\}=0$. Independence then
gives $\bbE\{H(U)q(Y)\}=0$ for every polynomial $q$ of degree at most
two. In particular the change in the population direct log likelihood is
\[
 Q_\delta(\theta)-Q_0(\theta)=\delta\bbE\!\left[H(U)
 \{\log f(Y;\theta)-\log f_C(C;\theta)\}\right]=0
\]
for every Gaussian candidate. Under A0 the true Gaussian law maximizes
each pattern's criterion and the complete-case term identifies it; the
perturbation therefore leaves its unique maximizer unchanged. Yet the
pattern-$1000$ perturbation averages to zero given $C$, so the
normalization error is $F(Y)=\delta H(U)$, which is not almost surely
zero. This example separates failure of CIM from bias of the fitted
Gaussian parameters.

\begin{corollary}[Identification for four-variable Gaussian alternatives]
\label{prop:gaussian-separated-alternatives}
For B1 and B2 with $\gamma>0$, and for the preceding example with
$0<\delta\leq1/4$, no
CIM law with a nondegenerate four-variate Gaussian complete-data
marginal has the same observed-data distribution. Consequently, a
characteristic-kernel discrepancy is positive at every Gaussian
candidate.
\end{corollary}

\begin{proof}
The coordinate $Y_4$ is observed only in pattern $1111$. For B1 and
B2, write $v=Y_4$, $X=(C,Y_2,Y_3)$, and let
$\psi_1(v)=1-2\expit(2v)$ and $\psi_2(v)=1-2\expit\{2(v^2-1)\}$, respectively.
The complete-case probability is
\[
\pi_{1111}(X,v)=
\frac{q_{1111}(C)}{A_j(X)+q_{1000}(C)\exp\{\gamma\psi_j(v)\}},
\qquad j=1,2,
\]
where
\[
A_j(X)=q_{1111}(C)+q_{1100}(C)\exp\{\gamma\psi_j(Y_3)\}
                    +q_{1010}(C)\exp\{\gamma\psi_j(Y_2)\}.
\]
Since $|\psi_j|\leq1$, the denominator is at most $e^\gamma$ and
$\pi_{1111}\geq0.05e^{-\gamma}$. For every fixed $X$, the
complete-case probability is nonconstant in $v$. All incomplete-pattern
probabilities conditional on their observed coordinates omit $Y_4$.
Their sum therefore cannot cancel this nonconstant dependence in the
CIM identity on the positive Gaussian product support. Thus B1 and B2
fail CIM. Their complete-case lower bound satisfies the condition of
Theorem~\ref{thm:gaussian-tail-identification}, which excludes
all Gaussian--CIM explanations. B1 is also monotone in $Y_4$ and
satisfies the main paper's conditional-shape theorem, but B2 is not
monotone and is not justified by that condition.

For the preceding example, non-CIM and the complete-case lower bound
were shown above. The same Gaussian-tail theorem therefore excludes
every Gaussian--CIM explanation.
Alternatively, all three alternatives satisfy
Proposition~\ref{prop:conditional-ratio-separation} with $V=Y_4$ and
$X=(C,Y_2,Y_3)$: their complete-case selection probabilities are
bounded away from zero and nonconstant in $Y_4$. Domination holds
at every nondegenerate Gaussian candidate because all densities are
positive. The argument in Remark~\ref{rem:observational-equivalence-power} and the characteristic
property give positive candidate discrepancy.
\end{proof}

This is a population separation result. Consistency of the fitted test
also requires convergence to a regular fixed-alternative limit, as in
Theorem~\ref{prop:rkhs-limit} of the main paper.

\subsection{Further shape criteria and pattern interactions}
\label{supp:shape-generalization}

The proof of Theorem~\ref{thm:conditional-shape-identification}{}
of the main paper uses, for Gaussian conditionals, that a
candidate-to-true density ratio cannot be both bounded and nonconstant
monotone and, for Student $t$ conditionals, that
$f(t\mid x;\theta)/f(-t\mid x;\theta)\to1$ as $t\to\infty$; the same
Bayes argument therefore extends to any conditional family with either
property, including univariate exponential-power families with exponent
greater than one and finite mixtures of Student $t$ conditionals;
Proposition~\ref{prop:positive-tail-limit} dispenses with monotonicity
under the tail condition \eqref{eq:shape-plateau-main}; for Gaussian
conditionals, a positive tail limit alone suffices.
Fixed coordinatewise bijections preserve CIM and conditional density
ratios, so the Gaussian case covers, for example, lognormal conditional
models through $\log V$.

The restriction on the candidate family is substantive. For example,
with $\phi$ denoting the standard normal density, let $f_0(v)=\{\phi(v-1)+\phi(v+1)\}/2$ and
$f_1(v)=3\phi(v-1)/4+\phi(v+1)/4$. Their ratio is
$1/2+\expit(2v)$. With complete-or-missing observations, selection
$\pi_{\mathbf1}(v)=1/4+\expit(2v)/2$ under $f_0$ gives the same observed law
as MCAR with probability $1/2$ under $f_1$. Hence bounded monotone
selection alone does not separate general Gaussian mixtures.

\begin{theorem}[Affine-logistic identification for general patterns]
\label{thm:logit-patterns}
Let $|\mathcal R|\geq2$, include the complete pattern $\mathbf1$, and let the true
and candidate complete-data distributions be nondegenerate multivariate
Student $t$ distributions on $\R^p$, with arbitrary finite positive
degrees of freedom. Suppose the true pattern probabilities have the form
\begin{equation}
 \bbP(R=r\mid Y=y)
 =\frac{\exp(\alpha_r+\beta_r^{\trans}y)}
 {\sum_{s\in\mathcal R}\exp(\alpha_s+\beta_s^{\trans}y)}.
 \label{eq:softmax}
\end{equation}
Let $J=\{j:\beta_{rj}\ne\beta_{\mathbf1,j}\text{ for some }r\in\mathcal R\}$.
If
\begin{equation}
 \text{for every }r\in\mathcal R\setminus\{\mathbf1\},\quad
 \text{there exists }j\in J\text{ with }r_j=0,
 \label{eq:coverage}
\end{equation}
then no CIM law with a Student $t$ marginal has the same observed-data
law as the true law. The comparison CIM mechanism is unrestricted.
The result also holds for finite mixtures of multivariate $t$
distributions.
\end{theorem}
\begin{proof}
For a candidate CIM law, write $\pi_{r,\theta}$ for its observed-pattern
probabilities. The complete-pattern probability required by
observed-data equivalence is
\[
 \pi_{\mathbf1,\theta}(y)=\pi_{\mathbf1}(y)\frac{f(y;\theta_0)}{f(y;\theta)}.
\]
For every $j\in J$, there is a choice of sign such that, holding
$y_{-j}$ fixed and taking $y_j$ to that signed infinity, the denominator
of \eqref{eq:softmax} relative to its complete-pattern numerator has an
exponentially increasing term. Thus $\pi_{\mathbf1}(y)$ decays exponentially.
A ratio of multivariate $t$ densities along a coordinate line has only
polynomial growth. The same is true for finite $t$ mixtures, whose
slowest-decaying components determine the power on that line.
Therefore $\pi_{\mathbf1,\theta}(y)$ tends to zero in the selected tail for
every $j\in J$ and every fixed $y_{-j}$.

Under CIM it must also satisfy
\[
 \pi_{\mathbf1,\theta}(y)=1-\sum_{r\in\mathcal R\setminus\{\mathbf1\}}
                  \pi_{r,\theta}(y^{(r)}).
\]
By \eqref{eq:coverage}, each term on the right omits at least one
coordinate of $J$. The alternating rectangular difference of this
identity over all coordinates in $J$ is consequently zero. For
$J=\{j,l\}$, for instance, writing $y^{j}$, $y^{l}$, and $y^{jl}$ for
the points with the indicated coordinates replaced by their alternate
values, the identity reads
\[
\pi_{\mathbf1,\theta}(y)
-\pi_{\mathbf1,\theta}(y^{j})
-\pi_{\mathbf1,\theta}(y^{l})
+\pi_{\mathbf1,\theta}(y^{jl})=0,
\]
because each right-hand term omits $y_j$ or $y_l$ and cancels within the
corresponding difference. The
complete-pattern ratio is continuous, so this rectangular identity,
initially almost everywhere, holds everywhere. In that identity,
sequentially send the alternate coordinate values to their selected
signed infinities. At each step all terms using that alternate value
vanish. Only $\pi_{\mathbf1,\theta}(y)$ at the base point remains, forcing
it to be zero. This contradicts its strict positivity.
\end{proof}

For a Gaussian complete-data family, the same affine-logistic
coverage condition also excludes a CIM representation. The true
complete-case probability itself tends to zero in the selected tail
of every coordinate of $J$. The rectangular-difference argument in the
proof therefore shows that the true law is non-CIM. The Gaussian tail
condition in Theorem~\ref{thm:gaussian-tail-identification}{} of the main paper holds for every affine-logistic
mechanism, and that theorem excludes all Gaussian--CIM comparison laws.

The exclusive-coordinate example with patterns $110$ and $011$ in
Section~\ref{supp:conditional-identification} does not require a complete
pattern. In contrast,
Theorem~\ref{thm:logit-patterns} uses the complete pattern but replaces
exclusivity by the coordinate-coverage condition. These are two
complementary routes to identification.

\subsection{Fixed-template Gaussian mixtures}
\label{supp:gaussian-mixture-identification}

The two arguments in main-paper Section~\ref{sec:identification} extend to finite Gaussian
mixtures with different additional restrictions. The complete-case
argument uses a fixed mixture template, whereas the exclusive-coordinate
argument uses a probit observation probability.

\label{supp:fixed-template-mixture-identification}

The Gaussian complete-case tail argument extends to a restricted
Gaussian mixture family in which the mixing weights and relative
component locations are fixed.

\begin{theorem}[Fixed-template Gaussian mixture identification]
\label{thm:fixed-template-mixture-identification}
Fix $K<\infty$, weights $w_k>0$ with $\sum_{k=1}^K w_k=1$, and vectors
$b_k\in\R^p$. Consider the complete-data family
\[
\mathcal M=\left\{f_{\mu,\Sigma}(y)
 =\sum_{k=1}^K w_k\phi_p(y;\mu+b_k,\Sigma):
 \mu\in\R^p,\ \Sigma\succ0\right\},
\]
where $\phi_p(\,\cdot\,;a,S)$ denotes a nondegenerate Gaussian density.
The template $K,(w_k,b_k)_{k=1}^K$ is common to all laws under comparison.
Include the complete pattern $\mathbf1=(1,\ldots,1)$, and suppose the
true law has marginal $f_{\mu_0,\Sigma_0}\in\mathcal M$ and, for every
$\epsilon>0$, admits $c_\epsilon>0$ and $M_\epsilon<\infty$ such that
\[
\pi_{\mathbf1}(y)=\bbP(R=\mathbf1\mid Y=y)
\ge c_\epsilon\exp(-\epsilon\|y\|^2)
\quad\text{for Lebesgue-a.e. }\|y\|\ge M_\epsilon.
\]
If a CIM law with complete-data marginal in $\mathcal M$ has the same
observed-data law, then its complete-data marginal equals
$f_{\mu_0,\Sigma_0}$ and the true law satisfies CIM. No tail restriction
is imposed on the comparison CIM law, and the remaining patterns are
arbitrary.
\end{theorem}

\begin{proof}
The argument is that of Theorem~\ref{thm:gaussian-tail-identification}
with two modifications. Write $f_0=f_{\mu_0,\Sigma_0}$, let
$f=f_{\mu,\Sigma}$ be the complete-data density of an observationally
equivalent CIM law, and write $f_{0,A}$ and $f_A$ for their marginals on
an observed coordinate set $A$ (both equal to one for $A=\varnothing$).
As in that proof, observational equivalence gives $\pi_{\mathbf1}f_0\le f$,
and since $\log f_{\mu,\Sigma}(y)=-\tfrac12y^{\trans}\Sigma^{-1}y+O(\|y\|)$
uniformly in direction (the mixture factor is a finite positive sum of
exponentials of affine functions of $y$), the tail condition forces
$D:=\Sigma-\Sigma_0\succeq0$.

First, because the template is common, $f_{0,A}=f_A$ if and only if
$\mu_A=\mu_{0,A}$ and $\Sigma_{AA}=\Sigma_{0,AA}$, since the template
contributes the same terms to the first two moments of both marginal
mixtures. Hence, with $\delta=\mu-\mu_0$, the direction $v=\delta+Dw$ can
be chosen exactly as in the proof of
Theorem~\ref{thm:gaussian-tail-identification}, with $w$ outside
finitely many proper affine subspaces, so that $v_A=0$ for every
unchanged marginal and
$v_A^{\trans}(\Sigma_{0,AA}^{-1}-\Sigma_{AA}^{-1})v_A>0$ for every
changed marginal with $D_{AA}\ne0$; for the latter,
$\log\{f_{0,A}(y_A+tv_A)/f_A(y_A+tv_A)\}
=-\tfrac12t^2v_A^{\trans}(\Sigma_{0,AA}^{-1}-\Sigma_{AA}^{-1})v_A+O(t)
\to-\infty$ as $t\to\infty$.

Second, for a changed marginal with $D_{AA}=0$, positive
semidefiniteness gives $D_{A,\{1,\ldots,p\}}=0$, so $v_A=\delta_A\ne0$
and $S:=\Sigma_{AA}=\Sigma_{0,AA}$. Factoring the common Gaussian density,
\[
f_A(x)=\phi_{|A|}(x;\mu_A,S)
 \sum_{k=1}^K w_k
 \exp\{b_{k,A}^{\trans}S^{-1}(x-\mu_A)
           -\tfrac12b_{k,A}^{\trans}S^{-1}b_{k,A}\},
\]
the mixture sums of $f_{0,A}$ and $f_A$ differ termwise by the fixed
factors $\exp(b_{k,A}^{\trans}S^{-1}\delta_A)$, so their ratio is
bounded above and below uniformly in $x$, and the Gaussian factors give
$\log\{f_{0,A}(y_A+t\delta_A)/f_A(y_A+t\delta_A)\}
=-t\delta_A^{\trans}S^{-1}\delta_A+O(1)\to-\infty$.

Thus along $y+mv$ every changed pattern marginal has $f_{0,A}/f_A\to0$
and every unchanged one is constant, and the remainder of the proof of
Theorem~\ref{thm:gaussian-tail-identification} applies verbatim:
all mixture marginals are strictly positive, so the comparison CIM
normalization and the identities $\pi'_r=\pi_rf_{0,A_r}/f_{A_r}$ hold
outside a Lebesgue-null set and along $y+mv$ for every $m$; letting
$m\to\infty$ shows that the probabilities of the unchanged patterns sum
to one, which contradicts $\rho_{\mathbf1}>0$ when the complete-case
marginal is changed. Hence $f=f_0$, $\pi'_r=\pi_r$ almost everywhere,
and the true law satisfies CIM.
\end{proof}

A uniform lower bound $\pi_{\mathbf1}(y)\ge c>0$ is a simple sufficient special
case. The theorem concerns a common fixed template, allowing only a
common location and common within-component covariance to vary.
Fixing a template at an estimate obtained from the same observed data
does not establish identification within the unrestricted mixture
family.

\subsection{Finite Gaussian mixtures and probit observation probabilities}
\label{supp:mixture-probit-identification}

The exclusive-coordinate argument extends to finite Gaussian mixtures
when the observation probability has a Probit form. Unlike the
Gaussian result, nonconstant monotonicity alone is insufficient for
this extension.

\begin{theorem}[Probit identification for finite Gaussian mixtures]
\label{thm:mixture-probit-separation}
Let $Y=(X,V,W)$, where $V$ is scalar, and suppose that
$Y^{(r)}=(X,V)$ for an observation pattern $r$. Assume that $V$
is unobserved in every other pattern. Consider a class $\mathcal P$ of
joint laws for $(Y,R)$ whose complete-data densities are strictly
positive, almost everywhere, on the same Cartesian support
$\mathcal X\times\mathbb R\times\mathcal W$. Under every law in
$\mathcal P$, suppose that the conditional distribution of $V\mid X=x$
is a finite mixture of nondegenerate Gaussian distributions with
positive weights for almost every $x$. The number of components and
their parameters may be unknown and may differ between laws.

For a law $P_{Y,R}\in\mathcal P$, suppose that, on a measurable set
$E$ with $P_X(E)>0$,
\[
\pi_r(x,v)
:=\bbP(R=r\mid X=x,V=v)
=\Phi\{\alpha(x)+\beta(x)v\}
\]
for almost every $(x,v)$, where $\alpha(x)$ and $\beta(x)$ are finite
and $\beta(x)\ne0$ for almost every $x\in E$. Here $\Phi$ denotes the
standard Gaussian distribution function. Then no CIM law in
$\mathcal P$ has the same observed-data law as $P_{Y,R}$.
\end{theorem}

\begin{proof}
Suppose that a comparison law $P'_{Y,R}\in\mathcal P$ satisfies CIM
and has the same observed-data law as $P_{Y,R}$. Its normalization
condition gives
\[
\pi'_r(X,V)
=1-\sum_{s\in\mathcal R\setminus\{r\}}\pi'_s(Y^{(s)}).
\]
The right-hand side does not depend on $V$. Positivity on the
Cartesian support and Fubini's theorem imply that
$\pi'_r(x,v)=c(x)$ for almost every $(x,v)$, for some measurable
$c$. Consequently, wherever this pattern has positive conditional
probability,
\[
\mathcal L_{P'}(V\mid X=x,R=r)
=\mathcal L_{P'}(V\mid X=x).
\]

Write $f(v\mid x)$ and $f'(v\mid x)$ for the true and comparison
conditional densities. On $E$, the Probit probability is strictly
positive, so $\bbP(R=r\mid X=x)>0$ for almost every $x$.
The common support and equality of the observed-data laws therefore
give, for almost every $x\in E$,
\[
f'(v\mid x)
=\frac{f(v\mid x)\Phi\{\alpha(x)+\beta(x)v\}}
       {\bbP(R=r\mid X=x)}
\quad\text{for almost every }v.
\]
Both sides have continuous versions as functions of $v$, so their
equality holds for every $v$. Choose an $x\in E$ at which these
properties hold, and suppress its dependence in the following
tail calculation.

If $\beta>0$, consider $v\to-\infty$; if $\beta<0$, consider
$v\to+\infty$. In either selected tail, a finite Gaussian mixture
has an asymptotic form
\[
f(v\mid x)=C\exp(-A v^2+Bv)\{1+o(1)\},\qquad A,C>0.
\]
Indeed, the components with largest variance dominate; among them,
the component with the most extreme mean in that tail dominates,
after identical components have been combined. Mills' ratio yields
\[
\Phi(\alpha+\beta v)
=\frac{\exp\{-(\alpha+\beta v)^2/2\}}
       {\sqrt{2\pi}\,|\beta|\,|v|}\{1+o(1)\}.
\]
Thus the selected conditional density has the form
\[
\frac{f(v\mid x)\Phi(\alpha+\beta v)}
     {\bbP(R=r\mid X=x)}
=\widetilde C\,|v|^{-1}
  \exp\{-(A+\beta^2/2)v^2+(B-\alpha\beta)v\}\{1+o(1)\},
\qquad \widetilde C>0.
\]
In contrast, the comparison finite Gaussian mixture satisfies
\[
f'(v\mid x)=C'\exp(-A'v^2+B'v)\{1+o(1)\},\qquad A',C'>0.
\]
Equality of the densities first forces
$A'=A+\beta^2/2$ by comparison of the quadratic terms, and then
$B'=B-\alpha\beta$ by comparison of the linear terms. With these
coefficients equal, the ratio of the selected density to
$f'(v\mid x)$ tends to zero because of the factor $|v|^{-1}$,
contradicting equality. Hence the proposed CIM comparison law
cannot exist.
\end{proof}

The comparison observation probabilities in
Theorem~\ref{thm:mixture-probit-separation} need not have a probit form.
The result separates the specified alternatives from every CIM candidate;
it does not identify all mixture parameters or CIM status throughout an
unrestricted mixture model. The conditional mixture assumption holds
for every finite multivariate Gaussian mixture. Related parameter
identification for a scalar Gaussian-mixture response with a specified
probit missingness model is established by \citet{MiaoDingGeng2016}; the
present result instead uses an exclusive observation pattern and imposes
no parametric restriction on the comparison CIM mechanism.

For example, with $Y=(Y_1,Y_2,Y_3)$ a finite Gaussian mixture and only
patterns $110$ and $011$, the selection probability
$\bbP(R=110\mid Y)=\Phi(\alpha+\beta Y_1)$ satisfies the theorem for
$\beta\ne0$, taking $V=Y_1$ and $X=Y_2$. No complete cases are needed.
For $\beta=0$ the mechanism is MCAR. Thus CIM status is identified within
this specified mechanism class, even though the theorem does not claim
identification of the complete-data mixture parameters.

Uniform positivity and nonconstant monotonicity alone are insufficient
for an unrestricted Gaussian mixture. The mixture-reweighting
counterexample in Section~\ref{supp:shape-generalization} has a
complete-case probability between $1/4$ and $3/4$ and a non-CIM true law,
yet admits an observationally equivalent MCAR mixture law. It explains
why the template restriction in
Theorem~\ref{thm:fixed-template-mixture-identification} and the probit
shape condition in Theorem~\ref{thm:mixture-probit-separation} cannot
simply be omitted.

\FloatBarrier

\subsection{Nonidentification within a Gaussian family}
\begin{remark}[Fitted compatibility and identification]
\label{rem:observational-equivalence-power}
Take the non-CIM law $Q_{Y,R}$ on the left of
Figure~\ref{fig:counterpart-schematic}, with $Q_Y=P_{Y,\theta_0}$.
Completion at $\theta_0$ gives its counterpart $(Q_{Y,R})^{\mathrm{cp}}$,
whose marginal differs from $Q_Y$. Completion at any candidate parameter
preserves the observed law $Q_O$ by construction. Suppose each
observed-pattern law is dominated by the corresponding marginal of
$P_{Y,\theta^*}$. If completion at the fitted limit $\theta^*$ satisfies $\widetilde P_{Y,\theta^*}=P_{Y,\theta^*}\neq Q_Y$, then, with $\rho_r$ and $f_r(\cdot\mid R=r)$ taken from $Q_O$, its conditional pattern probabilities are
\[
\widetilde P_{Y,R,\theta^*}(R=r\mid Y=y)
=\frac{\rho_r f_r(y^{(r)}\mid R=r)}{f_r(y^{(r)};\theta^*)},
\qquad r\in\mathcal R,
\]
$P_{Y,\theta^*}$-almost surely. The ratio follows by dividing the
conditional-completion joint density by its model marginal and depends
only on $y^{(r)}$. Thus $\widetilde P_{Y,R,\theta^*}$ is MAR, hence CIM,
and explains the same observed-data distribution as $Q_{Y,R}$. Zero
fitted discrepancy can therefore correspond to a CIM explanation even when the generating law is non-CIM.
Because the two laws induce the same observed-data distribution, every
observed-data test has the same rejection probability under both and
therefore cannot distinguish them.
\end{remark}

\begin{example}[Failure of identification of CIM]
\label{ex:gaussian-observational-equivalence}
Let $\mathcal M$ be the family of nondegenerate bivariate Gaussian
distributions. Under the true law $P_{Y,R}$, take
$P_Y=P_{Y,\theta_0}=N_2(0,I_2)$ and use patterns $11,10,01,00$.
Define the missing-data mechanism by
\[
\begin{aligned}
\bbP(R=11\mid Y=y)&=0.2\exp(-(y_1^2+y_2^2)/2),\\
\bbP(R=10\mid Y=y)&=0.1\sqrt{2}\exp(-y_1^2/2),\\
\bbP(R=01\mid Y=y)&=0.1\sqrt{2}\exp(-y_2^2/2),\\
\bbP(R=00\mid Y=y)&=1-\sum_{r\in\{11,10,01\}}
\bbP(R=r\mid Y=y).
\end{aligned}
\]
These probabilities are valid because their first three terms sum to at
most $0.2+0.2\sqrt{2}<1$. For $P_{Y,\theta'}=N_2(0, I_2/2)$,
integrating out the missing coordinates gives
\[
\bbP(R=r,Y^{(r)}\in du)=0.1 f_r(u;\theta')\,du,
\quad r\in\{11,10,01\},
\qquad \bbP(R=00)=0.7.
\]
Thus $P_O$ is exactly the observed-data law generated by
$P_{Y,\theta'}$ and an MCAR mechanism with pattern probabilities
$(0.1,0.1,0.1,0.7)$. Under $P_{Y,R}$,
$\pi_{00}=0.7$, while
$\pi_{11}$, $\pi_{10}$, and $\pi_{01}$ are the first three displayed
probabilities. Hence
\[
F(0,0)=0.2+0.2\sqrt{2}+0.7-1=0.2\sqrt{2}-0.1>0.
\]
Continuity gives CIM failure on a set of positive $P_Y$-probability.
Thus a non-CIM law and an MCAR law, with different complete-data
marginals in the same Gaussian family, have exactly the same
observed-data distribution. This violates
\eqref{eq:cim-membership-identification}.
\end{example}

\section{Simulation details and supplementary results}\label{supp:revised-protocol}

\subsection{Corrected-multiplier calibration}
\label{supp:auxiliary-multiplier-component}
Table~\ref{tab:retained-calibration} reports both corrected multiplier
calibrations with $B=1999$ on 500 datasets per null cell. The pooled
A1/A2 rows combine independent sets of simulations estimating the same
rejection probability. The two multiplier calibrations use the same
estimated contributions, so their agreement does not provide independent
evidence of size accuracy. For A1 at $n=2000$, both multiplier calibrations
give a rejection rate of $8.6\%$ (43/500; individual 95\% Monte Carlo
interval $6.3$--$11.4\%$), above the nominal $5\%$ level.

\begin{table}[ht]
\centering\tablebodysetup
\caption{Corrected-multiplier calibration under the Gaussian joint null: rejection at level $0.05$ and Monte Carlo uncertainty.}
\label{tab:retained-calibration}
\begin{tabular}{@{}lrlrrrl@{}}
\toprule
Case & $n$ & Calibration & Rejections & Rate & MC SE & Individual 95\% MC interval \\
\midrule
A0 & 500 & MULT\_G & 24/500 & 0.048 & 0.0096 & $[0.0310,0.0706]$ \\
A0 & 500 & MULT\_R & 29/500 & 0.058 & 0.0105 & $[0.0392,0.0822]$ \\
A0 & 2000 & MULT\_G & 25/500 & 0.050 & 0.0097 & $[0.0326,0.0729]$ \\
A0 & 2000 & MULT\_R & 27/500 & 0.054 & 0.0101 & $[0.0359,0.0776]$ \\
A1 & 500 & MULT\_G & 25/500 & 0.050 & 0.0097 & $[0.0326,0.0729]$ \\
A1 & 500 & MULT\_R & 31/500 & 0.062 & 0.0108 & $[0.0425,0.0869]$ \\
A1 & 2000 & MULT\_G & 43/500 & 0.086 & 0.0125 & $[0.0629,0.1141]$ \\
A1 & 2000 & MULT\_R & 43/500 & 0.086 & 0.0125 & $[0.0629,0.1141]$ \\
A2 & 500 & MULT\_G & 21/500 & 0.042 & 0.0090 & $[0.0262,0.0635]$ \\
A2 & 500 & MULT\_R & 23/500 & 0.046 & 0.0094 & $[0.0294,0.0682]$ \\
A2 & 2000 & MULT\_G & 24/500 & 0.048 & 0.0096 & $[0.0310,0.0706]$ \\
A2 & 2000 & MULT\_R & 25/500 & 0.050 & 0.0097 & $[0.0326,0.0729]$ \\
\addlinespace[3pt]
A1/A2 & 500 & MULT\_G & 46/1000 & 0.046 & 0.0066 & $[0.0339,0.0609]$ \\
A1/A2 & 500 & MULT\_R & 54/1000 & 0.054 & 0.0071 & $[0.0408,0.0699]$ \\
A1/A2 & 2000 & MULT\_G & 67/1000 & 0.067 & 0.0079 & $[0.0523,0.0843]$ \\
A1/A2 & 2000 & MULT\_R & 68/1000 & 0.068 & 0.0080 & $[0.0532,0.0854]$ \\
\bottomrule
\end{tabular}
\par\smallskip
\begin{minipage}{0.98\linewidth}\footnotesize
Each unpooled row uses 500 independent datasets; pooled A1/A2 rows
combine the two sets of 500 datasets from the same observed-data law.
Each calibration uses $B=1999$ multiplier draws per dataset.
Intervals are individual 95\% Clopper--Pearson intervals for the
rejection probability. A0--A2 satisfy CIM.
\end{minipage}
\end{table}
\FloatBarrier

\subsection{Generating mechanisms}\label{supp:numerical-design}\label{supp:moved-simulation-design}\label{supp:gaussian-baseline}

For the baseline mechanisms, generate $Y=(Y_1,Y_2,Y_3,Y_4)\sim N_4(0,\Sigma)$ with
$\Sigma_{jk}=0.5^{|j-k|}$, and write $C=Y_1$ for the always-observed
covariate. The observation patterns are $1111$, $1100$, $1010$, and
$1000$. A0 assigns each pattern probability $1/4$. For A1, define the
MAR probabilities
\[
q_r(C)=0.05+0.8\frac{\exp(a_r C)}{\sum_s\exp(a_s C)},
\qquad a=(0,0.5,-0.5,0.25),
\]
in the stated pattern order. For A2, change only
\begin{align*}
p_{1100}&=q_{1100}(C)+b(C)\{2\expit(2U)-1\},\\
p_{1010}&=q_{1010}(C)-b(C)\{2\expit(2U)-1\},
\end{align*}
where $b(C)=0.8\min\{q_{1100}(C),q_{1010}(C)\}$ and
$U=(Y_4-0.5Y_3)/\sqrt{0.75}$. 
Writing $H=2\expit(2U)-1$, the generating covariance gives
\[
\Cov\{Y_4-0.5Y_3,(C,Y_2,Y_3)\}=(0,0,0),
\qquad \Var(Y_4-0.5Y_3)=0.75.
\]
Joint Gaussianity implies that $U$ is standard normal and independent
of $(C,Y_2,Y_3)$. Its symmetry therefore gives
$\bbE(H\mid C,Y_2,Y_3)=0$. Since $|H|\leq1$, each affected probability
is at least $0.2q_r(C)>0$, and the four probabilities still sum to one.

Both affected patterns omit $Y_4$. Conditional averaging over their
missing components eliminates $b(C)H$, so every pattern subdistribution
equals its A1 counterpart. In particular,
$\pi_r(Y^{(r)})=q_r(C)$ for every pattern under A2, and their sum is
one. Thus A2 satisfies CIM while its probabilities depend on the
unobserved $Y_4$; it is not MAR. The complete-data marginal is also
unchanged. Any fixed observed-data estimator consequently has exactly
the same sampling distribution under A1 and A2. The transfer results
in Theorem~\ref{thm:estimation-transfer} and
Corollary~\ref{cor:efficiency-transfer} additionally identify the
same attainable efficiency bound.
The Monte Carlo design, the Gaussian fit, the targets and the kernel are
those of Section~\ref{sec:simulation} of the main paper.

For the non-CIM mechanisms B1 and B2, retain the Gaussian complete-data distribution, baseline $q_r(C)$,
and observation patterns of A, and use
\[
p_r(Y)=\frac{q_r(C)\exp\{\gamma h_r(Y)\}}
{\sum_s q_s(C)\exp\{\gamma h_s(Y)\}},\qquad
\gamma\in\{0.5,1\},
\]
where the pattern-specific tilts in the stated order are
\[
\begin{aligned}
\mathrm{B1}:\quad h&=\bigl(0,1-2\expit(2Y_3),1-2\expit(2Y_2),1-2\expit(2Y_4)\bigr),\\
\mathrm{B2}:\quad h&=\bigl(0,1-2\expit\{2(Y_3^2-1)\},1-2\expit\{2(Y_2^2-1)\},\\
&\hspace{3.5em}1-2\expit\{2(Y_4^2-1)\}\bigr).
\end{aligned}
\]
Section~\ref{supp:original-gaussian-separation} verifies that both
selections fail CIM and are separated from every Gaussian--CIM
explanation.

The two MCAR full-model misspecification controls each have mean zero
and covariance $\Sigma$. For the Gaussian mixture, generate
$Y=W+0.6Se_4$, where $W\sim N_4(0,\Sigma-0.36e_4e_4^{\trans})$
and $S\in\{-1,1\}$ is independently equiprobable. For the $t_{10}$
control, generate $Y=X\sqrt{8/Q}$, with $X\sim N_4(0,\Sigma)$ and
independent $Q\sim\chi^2_{10}$; its scale matrix is $0.8\Sigma$.
Both assign probability $1/4$ to each pattern.

To check the size of the estimated-coordinate rule, separately from the fixed coordinates in Experiments A and B, we
applied the paired-mean and paired-SD rule with global centring to
$500$ independent samples of size $250$ per law, using $199$ row
bootstrap draws per sample. Let $(Z_1,Z_2)$ be standard bivariate
Gaussian with correlation $0.4$ and set $Y=(5+0.5Z_1,10.5+2Z_2)$.
Under MCAR the probabilities for $11,10,01,00$ were
$(0.696,0.152,0.120,0.032)$. Under MAR, keep $q_{00}=0.032$,
add $0.115\tanh Z_1$ and $0.090\tanh Z_2$ to $q_{10}$ and $q_{01}$,
respectively, and choose $q_{11}$ by subtraction. At nominal $5\%$,
the rejection rates were $26/500=5.2\%$ and $12/500=2.4\%$
(Monte Carlo SEs $0.99$ and $0.68$ percentage points).
The test was conservative in the second setting. Adding
$0.02\tanh\{(Z_2-0.4Z_1)/\sqrt{0.84}\}$ to $q_{10}$ and subtracting
it from $q_{00}$ gives a non-MAR CIM mechanism with the same observed
and full-data marginals: the perturbation has conditional mean zero
given $Z_1$. Thus the diagnostic has the same rejection probability
under these two mechanisms, by equality of their observed-data laws,
illustrating Theorem~\ref{thm:estimation-transfer}.

\subsection{Target inference in Experiment B}\label{supp:B-inference}
Table~\ref{tab:B-inference} summarizes estimation of the two targets in
the 18 settings of Experiment~B, using the same simulated datasets and
bootstrap refits as the main-paper diagnostic.

\begin{table}[htbp]
\centering\tablebodysetup
\setlength{\tabcolsep}{6pt}
\caption{Experiment B: target inference across all 18 settings.
Each row summarizes 500 independent Monte Carlo iterations. The targets
are $\bbE(Y_4)=0$ and $\Cov(Y_2,Y_3)=0.5$. Coverage of the 95\%
percentile intervals is reported as a percentage, with its Monte Carlo
standard error in percentage points in parentheses. Intervals use the
same $B=499$ whole-row bootstrap refits as the diagnostic.}
\label{tab:B-inference}
\begin{tabular}{llrrrr}
\toprule
& & \multicolumn{2}{c}{Mean $Y_4$}
& \multicolumn{2}{c}{$\Cov(Y_2,Y_3)$} \\
\cmidrule(lr){3-4}\cmidrule(lr){5-6}
Law & $n$ & Bias / RMSE & Coverage & Bias / RMSE & Coverage \\
\midrule
B1, $\gamma=0.5$ & 500 & 0.085 / 0.119 & 83.6 (1.7) & 0.009 / 0.079 & 95.2 (1.0) \\
B1, $\gamma=0.5$ & 1000 & 0.086 / 0.104 & 70.2 (2.0) & 0.009 / 0.056 & 93.0 (1.1) \\
B1, $\gamma=0.5$ & 2000 & 0.084 / 0.095 & 49.8 (2.2) & 0.006 / 0.041 & 94.0 (1.1) \\
B1, $\gamma=1$ & 500 & 0.170 / 0.192 & 47.4 (2.2) & 0.017 / 0.081 & 93.6 (1.1) \\
B1, $\gamma=1$ & 1000 & 0.174 / 0.185 & 18.6 (1.7) & 0.020 / 0.059 & 92.6 (1.2) \\
B1, $\gamma=1$ & 2000 & 0.172 / 0.177 & 2.4 (0.7) & 0.018 / 0.043 & 93.0 (1.1) \\
B2, $\gamma=0.5$ & 500 & 0.017 / 0.099 & 91.8 (1.2) & 0.053 / 0.101 & 89.0 (1.4) \\
B2, $\gamma=0.5$ & 1000 & -0.001 / 0.065 & 94.2 (1.0) & 0.048 / 0.076 & 86.8 (1.5) \\
B2, $\gamma=0.5$ & 2000 & 0.006 / 0.045 & 95.8 (0.9) & 0.055 / 0.070 & 73.6 (2.0) \\
B2, $\gamma=1$ & 500 & 0.011 / 0.109 & 92.8 (1.2) & 0.104 / 0.137 & 77.0 (1.9) \\
B2, $\gamma=1$ & 1000 & 0.011 / 0.071 & 95.6 (0.9) & 0.107 / 0.124 & 56.8 (2.2) \\
B2, $\gamma=1$ & 2000 & 0.008 / 0.051 & 95.2 (1.0) & 0.111 / 0.119 & 26.6 (2.0) \\
Mixture & 500 & 0.000 / 0.079 & 95.0 (1.0) & 0.008 / 0.079 & 94.0 (1.1) \\
Mixture & 1000 & -0.001 / 0.054 & 97.4 (0.7) & 0.001 / 0.054 & 94.8 (1.0) \\
Mixture & 2000 & 0.000 / 0.043 & 95.2 (1.0) & 0.000 / 0.039 & 95.4 (0.9) \\
$t_{10}$ & 500 & 0.002 / 0.079 & 95.2 (1.0) & -0.004 / 0.090 & 94.4 (1.0) \\
$t_{10}$ & 1000 & -0.001 / 0.060 & 93.4 (1.1) & 0.003 / 0.067 & 93.0 (1.1) \\
$t_{10}$ & 2000 & -0.001 / 0.042 & 94.2 (1.0) & 0.003 / 0.047 & 93.4 (1.1) \\
\bottomrule
\end{tabular}
\end{table}
\FloatBarrier

\section{SUPPORT2 application details}\label{supp:support2-checks}

\subsection{Cohorts and pattern summaries}
The public \texttt{support2csv.zip} file is distributed by the Vanderbilt
Biostatistics data repository \citep{HarrellSupportData}.
The cohort comprises all $3515$ records labelled ARF/MOSF w/Sepsis
in \texttt{dzgroup}, with pattern counts $(1962,293,1044,216)$ in
order $11,10,01,00$;
Table~\ref{tab:support2-patterns} gives original-scale summaries. Rows
with both laboratories missing have observed likelihood one and zero
centred embedding contribution but remain in $n$ and in the resampling.

\begin{table}[ht]
\centering\tablebodysetup
\caption{SUPPORT2 summaries by laboratory observation pattern.
Means and standard deviations are on the original measurement scales;
a dash denotes a measurement missing by definition of the pattern.}
\label{tab:support2-patterns}
\begin{tabular}{@{}lrrrrrr@{}}
\toprule
& & & \multicolumn{2}{c}{Albumin} & \multicolumn{2}{c}{Oxygenation}\\
\cmidrule(lr){4-5}\cmidrule(l){6-7}
Pattern & $n$ & Percent & Mean & SD & Mean & SD\\
\midrule
$11$ & 1962 & 55.82 & 2.722 & 0.729 & 223.3 & 107.2\\
$10$ & 293  & 8.34  & 3.153 & 0.761 & --- & ---\\
$01$ & 1044 & 29.70 & --- & --- & 224.4 & 101.4\\
$00$ & 216  & 6.15  & --- & --- & --- & ---\\
\bottomrule
\end{tabular}
\end{table}

\subsection{Primary bivariate analysis}
\label{supp:moved-support-qualification}
For $(\sqrt{\mathrm{alb}},\sqrt{\mathrm{pafi}})$ the paired sample means
are $(1.6348,14.527)$ and the sample SDs are $(0.2218,3.507)$, using
$m=1962$ and denominator $m-1$ for the SDs. Calculations use unrounded
values. In each of $1999$ whole-row bootstrap resamples, identify the
paired observations anew, recompute both means and SDs, and refit all
five Gaussian parameters. Compare the signed embeddings in the fixed
base-kernel RKHS with the respective original and resampled coordinate
maps, as in Section~\ref{supp:estimated-scales}. The first panel of
Table~\ref{tab:support2-tests} gives these results.

\begin{table}[ht]
\centering\tablebodysetup
\setlength{\tabcolsep}{3.5pt}
\caption{SUPPORT2 tests of joint Gaussian--CIM compatibility.
All analyses estimate measurement locations and scales. Panel A uses two
laboratories; panel B also estimates the covariate means and SDs used by
the product kernel. Each row uses $1999$ draws.}
\label{tab:support2-tests}
\begin{tabular*}{\textwidth}{@{\extracolsep{\fill}}llrrrr@{}}
\toprule
Specification & Calibration & $n$ & $T$ & $c_{0.95}$ & $p$ \\
\midrule
\multicolumn{6}{@{}l}{\textit{A. Bivariate: estimated means and SDs}}\\
Bivariate & Bootstrap & 3515 & 1.179 & 0.1405 & 0.0005 \\
 & Gaussian multiplier & 3515 & 1.179 & 0.1432 & 0.0005 \\
 & Rademacher multiplier & 3515 & 1.179 & 0.1428 & 0.0005 \\
\addlinespace[2pt]
\multicolumn{6}{@{}l}{\textit{B. Conditional sensitivity analyses: estimated means and SDs}}\\
4 labs, linear mean & Bootstrap & 3513 & 3.390 & 0.2834 & 0.0005 \\
6 labs, linear mean & Bootstrap & 3513 & 5.111 & 0.3590 & 0.0005 \\
4 labs, fixed-knot mean & Bootstrap & 3513 & 3.324 & 0.2635 & 0.0005 \\
\bottomrule
\end{tabular*}
\par\smallskip\begin{minipage}{\textwidth}\footnotesize
Panel A uses Algorithm~\ref{alg:cim-calibration} or the Gaussian
and Rademacher multipliers of Section~\ref{supp:multiplier} with
five Gaussian parameters, two paired means and two paired SDs. In panel B
each bootstrap draw refits the conditional Gaussian model and recomputes
the laboratory and covariate means and SDs.
The critical value $c_{0.95}$ uses type-7 quantiles.
\end{minipage}
\end{table}

\subsection{Original-scale target summaries}
\label{supp:support2-targets}
If $Z_j=(Y_j-\widehat a_j)/\widehat s_j$ and $(\mu,\Sigma)$ denote
the fitted Gaussian parameters for $Z$, the fitted mean of the original
laboratory $Y_j^2$ is
$(\widehat a_j+\widehat s_j\mu_j)^2+\widehat s_j^2\Sigma_{jj}$.
The fitted original-scale means are $2.778$ and $224.2$.
Applying the delta method to the observed-score sandwich covariance
gives standard errors $0.016$ and $1.916$ and 95\% intervals
$[2.747,2.809]$ and $[220.4,228.0]$. The complete-case means are
$2.722$ and $223.3$ from $1962$ patients; available-case means are
$2.778$ ($n=2255$) and $223.7$ ($n=3006$), respectively. These are
model-based summaries after rejection of joint compatibility, not
bias-corrected inference for the unknown complete-data means.

\subsection{Conditional specifications and observation patterns}
\label{supp:support2-conditional-details}\label{sec:support2-conditional}
The conditional cohort consists of the $3513$ patients with
$C=(\mathrm{age},\mathrm{aps},\mathrm{scoma},\mathrm{meanbp})$ jointly
observed. Two patients lacking a covariate are excluded (laboratory
patterns $0000$ and $0011$). We use the released APS variable, including
its imputed components.

Let $Y$ contain square-root albumin and oxygenation and log creatinine
and blood urea nitrogen (\texttt{crea}, \texttt{bun}); the six-laboratory
specification adds log bilirubin and white blood cell count
(\texttt{bili}, \texttt{wblc}). Write
$X_j=(Y_j-\widehat a_j)/\widehat s_j$. In these coordinates the primary
model is $X\mid C=c\sim N_4\{\Gamma^{\trans}z(c),\Sigma\}$, with
the analogous six-laboratory form. Use the paired-observed means and
SDs for the first two laboratories, and each laboratory's own observed
values for the remaining variables. The corresponding counts are
$(1962,1962,3495,1981,2521,3466)$; total observed counts are
$(2255,3006,3495,1981,2521,3466)$. In that order the original-sample means
are $(1.6348,14.527,0.3952,3.1990,0.2348,2.3795)$ and the SDs are
$(0.2218,3.507,0.7310,0.8025,1.172,0.7682)$. In the covariate kernel
each covariate is centred and divided by its SD over the $3513$ eligible
patients; the means are $(61.07,45.88,14.27,83.52)$ and the SDs are
$(17.63,19.59,23.45,29.57)$.
Each bootstrap sample recomputes every mean and SD from its corresponding
observed subset.

The regression design uses the fixed divisors $(10,10,10,20)$ as an
internal parameterization, separate from the kernel coordinates, so
covariate-kernel rescaling requires no transformation of the
coefficients. The linear mean uses an intercept and the four design columns.
With $d_j$ the $j$th fixed divisor, the cubic specification adds
$\{(c_j-\kappa)/d_j\}_+^3$ at each of three fixed knots per covariate.
No knot is reselected in any resample.
The knot locations are age $40,60,80$; APS $30,60,90$; coma score
$10,30,50$; and mean blood pressure $60,80,100$. The regression and
free-covariance parameter counts are $30$, $51$ and $78$ for the
four-laboratory linear, six-laboratory linear and four-laboratory
fixed-knot specifications. The model for $C$ remains unspecified
throughout. Table~\ref{tab:support2-conditional-pattern4} lists the 12
observed four-laboratory patterns; the six-laboratory cohort has 32
patterns among the same $3513$ patients, with $814$ complete records and
eight with all six laboratories missing, all retained.

\begin{table}[htbp]
\centering\tablebodysetup
\caption{Conditional four-laboratory cohort: all 12 observed patterns
in the order albumin, oxygenation, creatinine, blood urea nitrogen.
Both four-laboratory mean specifications use these same $3513$ patients.}
\label{tab:support2-conditional-pattern4}
\begin{tabular}{@{}lr@{\qquad\qquad}lr@{}}
\toprule
Pattern & $n$ & Pattern & $n$ \\
\midrule
$0000$ & 12 & $0110$ & 454 \\
$0001$ & 1 & $0111$ & 585 \\
$0010$ & 16 & $1010$ & 27 \\
$0011$ & 185 & $1011$ & 266 \\
$0100$ & 4 & $1110$ & 1019 \\
$0101$ & 1 & $1111$ & 943 \\
\midrule
\multicolumn{3}{r}{Total} & 3513 \\
\bottomrule
\end{tabular}
\end{table}

The product kernel applies the same mixture separately to the standardized
covariates and the laboratory vector $X$, then multiplies the two results:
\[
 k_0(u,v)=\frac13\sum_{h\in\{1,2,4\}}\exp\{-\|u-v\|^2/(2h^2)\}.
\]
In the following expression $k$ includes the covariate standardization,
while the conditioning variable $C$ is kept in its original regression
coordinates:
\begin{equation}
\xi_C(O;\theta)=
\bbE_\theta\{k((C,X),\cdot)\mid C,X^{(R)}\}
-\bbE_\theta\{k((C,X),\cdot)\mid C\},
\quad O=(C,R,X^{(R)}),
\label{eq:support2-conditional-contribution}
\end{equation}
whose second term is the individual's conditional model embedding
(Section~\ref{supp:conditional-reference-centering}). Rows with no
observed laboratory values contribute zero but remain in the sample.
Observed likelihood contributions are the conditional Gaussian marginal
densities of each patient's observed laboratory subset, and EM updates
the regression matrix and common covariance. Gaussian completion
supplies analytic laboratory-kernel expectations, multiplied by the
covariate-kernel factor; each original and bootstrap sample uses its own
fit and its own estimated means and SDs in the common base-kernel RKHS,
including the resample--original cross terms. Table~\ref{tab:support2-tests}
reports the results of the $1999$ whole-row refits for each model.

\subsection{Density reconstruction}
\label{supp:support2-density-details}\label{sec:support2-density}
\begin{figure}[ht]
\centering
\includegraphics[width=0.32\textwidth]{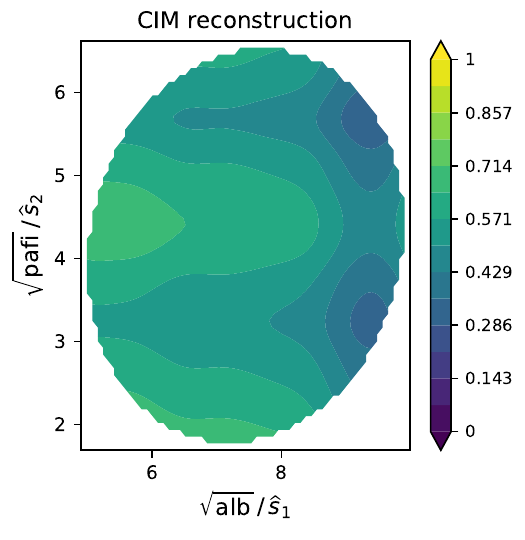}\hfill
\includegraphics[width=0.32\textwidth]{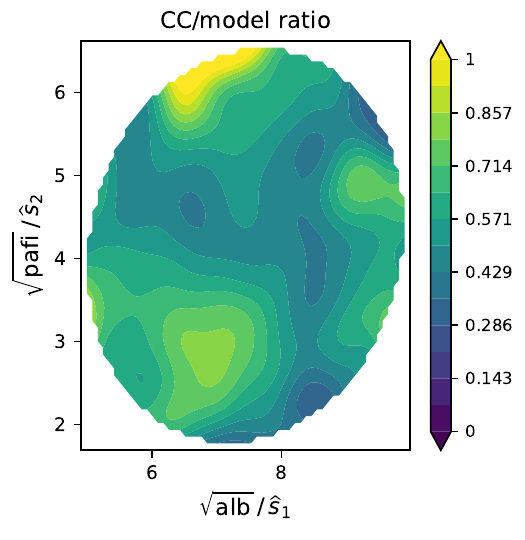}\hfill
\includegraphics[width=0.32\textwidth]{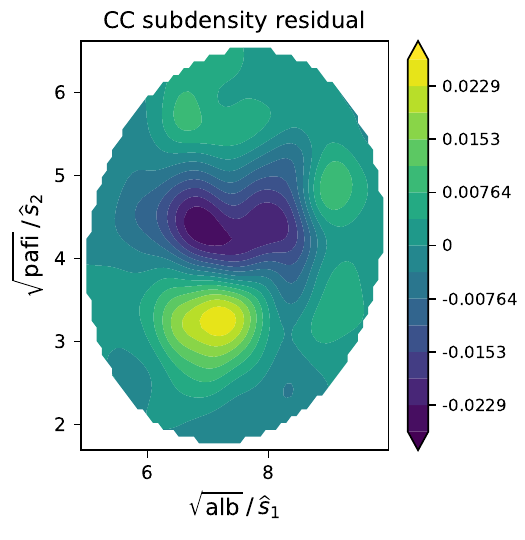}
\caption{Descriptive Gaussian--CIM density diagnostic.
Left: estimated right-hand side of
\eqref{eq:support2-additive-normalization}. Middle: estimated left-hand
side. Right: middle minus left, multiplied by $f_{11}(x;\widehat\theta)$.
The axes divide by the paired sample SDs without subtracting the means; the display shows the central fitted
95\% Gaussian ellipse.}
\label{fig:support2-density}
\end{figure}

For the density reconstruction, use the coordinates
\[x=(\sqrt{\mathrm{alb}}/\widehat s_1,\sqrt{\mathrm{pafi}}/\widehat s_2).\]
Use the density notation
of Section~\ref{sec:setup}, with $\mathcal R=\{11,10,01,00\}$.
Under joint Gaussian--CIM compatibility, \eqref{eq:pi} and
\eqref{eq:cim} imply
\begin{equation}
\frac{\rho_{11} f_{11}(x_1,x_2\mid R=11)}{f_{11}(x_1,x_2;\theta)}
=1-\frac{\rho_{10} f_{10}(x_1\mid R=10)}{f_{10}(x_1;\theta)}
  -\frac{\rho_{01} f_{01}(x_2\mid R=01)}{f_{01}(x_2;\theta)}-\rho_{00}.
\label{eq:support2-additive-normalization}
\end{equation}
Figure~\ref{fig:support2-density} compares the right side, reconstructed
from incomplete patterns, with the left side estimated from complete
cases. Their difference, multiplied by the fitted full density, locates
mismatches between the pattern-weighted densities. Pattern proportions
are empirical and within-pattern densities use kernel smoothing.
Computational details follow below.

The bivariate density display uses a $61\times61$ grid and
within-pattern Scott smoothing, transformed to the original-sample SD
coordinates. Ratios are invariant to this linear change of coordinates;
subdensities and their residuals include its Jacobian factor.
The pattern-weighted densities $\rho_r f_r(\cdot\mid R=r)$ integrate to $\rho_r$.
The displayed region is the central fitted 95\% Gaussian ellipse;
values outside $[0,1]$ are not truncated. The subdensity residual is
shown alongside ratios to reduce reliance on unstable tail divisions.

The residual is negative near intermediate albumin and oxygenation and
positive at lower and higher oxygenation, which is compatible with
non-Gaussian shape or oxygenation-dependent recording. Heavier-tailed
or mixture full models provide possible extensions of this analysis.

\FloatBarrier

\section{Change in childhood BMI in the Terneuzen Birth Cohort}\label{sec:tbc}
The Terneuzen Birth Cohort study examined the association between changes
in childhood body mass index (BMI) and adult overweight, identifying
ages 2 to 6 years as a period of particular interest \citep{DeKroon2010}.
We estimated the mean change in BMI standard deviation score (SDS) over
this period using the public \texttt{tbc} extract from the \texttt{mice}
R package (\url{https://amices.org/mice/reference/tbc.html}), which
contains 3,951 visits from 306 children. This exploratory analysis
used the following specification.

For each child we selected the visit closest to age 2 in $[1.5,2.5)$
years and the visit closest to age 6 in $[5.5,6.5)$ years. Ties were
resolved by earlier age, then by recorded occasion number (\texttt{occ}). Selection used
age alone: a missing score at the selected visit remained missing.
We used the recorded scores without interpolation. In age order, the
$11$, $10$, $01$ and $00$ patterns contained 91, 61, 27 and 127 children.
Most missing scores reflected the absence of a visit in the window:
149 of 154 at age 2 and 187 of 188 at age 6.

We fitted an unrestricted bivariate Gaussian by direct likelihood and
compared its estimates with those from the 91 complete pairs
(Table~\ref{tab:tbc-estimates}). The 88 children measured at only one
age contributed their observed marginal likelihoods. Children with
neither score contributed no information about the BMI parameters,
but all 306 children were retained when resampling.
Including measurements available at only one age changed the estimated mean difference
by $-0.014$ SDS and reduced its bootstrap standard error by $10\%$
in this sample.

\begin{table}[htbp]
\centering\tablebodysetup
\caption{BMI standard deviation scores near ages 2 and 6 in the
Terneuzen Birth Cohort. Standard errors in parentheses use the same
1,999 bootstrap samples of children for both methods.}
\label{tab:tbc-estimates}
\begin{tabular}{@{}lrr@{}}
\toprule
Target & Direct likelihood & Complete cases \\
\midrule
Mean change (age 6 minus age 2) & $-0.489$ $(0.100)$ & $-0.475$ $(0.111)$ \\
Mean near age 2 & $0.275$ $(0.079)$ & $0.274$ $(0.101)$ \\
Mean near age 6 & $-0.214$ $(0.085)$ & $-0.201$ $(0.098)$ \\
Correlation between ages & $0.389$ $(0.128)$ & $0.376$ $(0.125)$ \\
\bottomrule
\end{tabular}
\end{table}

For the diagnostic we used the equal-weight Gaussian kernel mixture
with bandwidths $1,2,4$. In the original sample and each bootstrap
refit, the kernel coordinates used the means and sample standard
deviations of the paired observations in that sample
(Section~\ref{supp:estimated-scales}). The observed statistic was
$T=0.06147$ and the 95\% critical value was $0.09888$; 470 of 1,999
centred bootstrap statistics were at least as large as $T$, giving
$\widehat p=0.2355$. The diagnostic did not reject joint Gaussian--CIM
compatibility.

\end{document}